\documentclass[twocolumn]{aastex62}
\usepackage{hyperref,astrojournals,amsmath,amssymb,mathtools,graphicx,txfonts,microtype,nicefrac,xspace,bm}
\usepackage{enumitem,kantlipsum}

\renewcommand*{\vec}[1]{\ensuremath{\boldsymbol{#1}}}
\let\oldhat\hat
\renewcommand*{\hat}[1]{\oldhat{\vec{#1}}}

\newcommand{\unitvectorslope}[1]{\ensuremath{%
\frac{\c{\oldhat{#1}}{z}}%
     {\sqrt{1-\c{\oldhat{#1}}{z}^2}}}}
\renewcommand{\c}[2]{\ensuremath{#1_{\text{#2}}}} 	
\newcommand{\ia}{\c{i}{a}\xspace}
\newcommand{\ib}{\c{i}{b}\xspace}
\newcommand{\ie}{\c{i}{e}\xspace}

\renewcommand*{\vec}[1]{\boldsymbol{#1}}

\hypersetup{%
  pdftitle={Apsidal Alignment},
  pdfauthor={\textcopyright\ authors},
  bookmarksopen=true,
  colorlinks=true,
  linkcolor=black,
  citecolor=black,
  urlcolor=black}

\begin{document}

\title{A Lopsided Outer Solar System?}

\shorttitle{Apsidal clustering} 

\shortauthors{Zderic et al.}
 
 \author[0000-0003-2961-4009]{Alexander Zderic}
\affiliation{JILA and Department of Astrophysical and Planetary Sciences, CU Boulder, Boulder, CO 80309, USA}
\email{alexander.zderic@colorado.edu}

\author[0000-0003-1971-1809]{Maria Tiongco}
\affiliation{JILA and Department of Astrophysical and Planetary Sciences, CU Boulder, Boulder, CO 80309, USA}

\author[0000-0002-0411-2742]{Angela Collier}
\affiliation{JILA and Department of Astrophysical and Planetary Sciences, CU Boulder, Boulder, CO 80309, USA}

\author[0000-0003-2527-4836]{Heather Wernke}
\affiliation{JILA and Department of Astrophysical and Planetary Sciences, CU Boulder, Boulder, CO 80309, USA}

\author[0000-0001-9261-0989]{Aleksey Generozov}
\affiliation{JILA and Department of Astrophysical and Planetary Sciences, CU Boulder, Boulder, CO 80309, USA}

\author[0000-0002-1119-5769]{Ann-Marie Madigan}
\affiliation{JILA and Department of Astrophysical and Planetary Sciences, CU Boulder, Boulder, CO 80309, USA}

\begin{abstract}
Axisymmetric disks of eccentric orbits in near-Keplerian potentials are unstable and undergo exponential growth in inclination. Recently, \citet{zderic2020a} showed that an idealized disk then saturates to a lopsided mode. Here we show, using $N$-body simulations, that this apsidal clustering also occurs in a primordial scattered disk in the outer solar system which includes the orbit-averaged gravitational influence of the giant planets. We explain the dynamics using \citet{lyndenbell1979}'s mechanism for bar formation in galaxies. 
We also show surface density and line of sight velocity plots at different times during the instability, highlighting the formation of concentric circles and spiral arms in velocity space. 
\end{abstract}

\keywords{celestial mechanics-minor planets, asteroids: general-planets and satellites: dynamical evolution and stability}

\section{Introduction}
\label{intro}

Something odd is going on in the outer solar system: distant bodies in orbit beyond Neptune appear clustered in argument of perihelion  \citep[$\omega$;][]{Trujillo2014} and longitude of perihelion  \citep[$\varpi$;][]{Batygin2016}. Some have extreme inclinations that cannot be generated in the standard model of solar system evolution \citep{Gladman2009, Chen2016, Becker2018, Kaib2019}, and others are ``detached'' in the sense that they have perhelia that lie far beyond the gravitational reach of the giant planets \citep[e.g.,][]{Brown04}. 
Observational biases have been carefully demonstrated in outer solar system surveys \citep{Shankman2017,Lawler2017,Kavelaars2020,Napier2021} but whether they can fully explain the anomalous orbital structure of  Trans-Neptunian objects (TNOs) remains a contentious issue \citep{Brown2017, Brown2019}.
If they do not, the outer solar system requires a new source of gravitational perturbation. One such source could be a planet far beyond the orbit of Neptune (for reviews see \citet{Batygin2019} and \citet{Trujillo2020}). We propose a different, internal source: the self-gravity of the bodies themselves.

The collective gravity of bodies on eccentric orbits in an axisymmetric near-Keplerian disk drives a dynamical instability \citep{Madigan2016,Madigan2018b}. This ``inclination instability'' exponentially grows the inclinations of orbits while decreasing their eccentricities, raising their perihelia and clustering their arguments of perihelion ($\omega$).
In a recent paper, \citet{Zderic2020b}, we showed that $\mathcal{O}(20)$ Earth masses are required for the instability to occur in a primordial scattered disk between $\sim10^2-10^3$ AU in the solar system under the orbit-averaged, gravitational influence of the giant planets at their current locations.
The instability can also generate a gap in perihelion at $\sim50 - 75$ AU, as observed in the outer solar system \citep{Kavelaars2020, Oldroyd2021}. 
The saturation timescale, that is, the time at which inclinations cease exponential growth, for the instability in a 20 Earth mass disk is far less than the age of the solar system. Therefore, to connect to the present-day outer solar system we need to understand the non-linear, saturated state of the instability. We are further motivated by the results of \citet{zderic2020a} where we discovered late-time apsidal clustering of orbits in the disk plane, albeit in simulations with highly idealized initial conditions. Here we show that the same late-time  clustering occurs in a primordial scattered disk between $\sim10^2-10^3$ AU in the solar system under the gravitational influence of the giant planets.
We essentially take the more realistic simulation conditions of \citet{Zderic2020b} and extend them past saturation to look for in-plane clustering.
We show that the apsidal clustering can be explained using \citet{lyndenbell1979}'s mechanism for bar formation in disk galaxies.

Our paper proceeds as follows: in \S\ref{sec:LB} we describe the Lynden-Bell mechanism for bar formation and show how it may be applied to near-Keplerian systems. In \S\ref{sec:methods} we describe our numerical methods and in \S\ref{sec:results} present our results. In  
 \S\ref{sec:observables} we show surface density and line of sight velocity plots of our simulations at different times, and we conclude in \S\ref{sec:conclusions}. 
\section{The Lynden-Bell Mechanism in near-Keplerian Systems}
\label{sec:LB}

In \citeyear{lyndenbell1979}, Donald Lynden-Bell described a mechanism by which bars may be formed in the centers of galaxies. 
We reproduce the basic argument here.

In a general galactic potential, a typical orbit is a rosette with an angle between $\pi$ and $2\pi$ linking consecutive apocenters.
If we view an orbit from rotating axes, we may choose the rotation speed $\nu_i$ such that the angle between apocenters will be $\pi$. The orbit will then be bisymmetric, like a centered oval or ellipse.
If $\nu$ is the mean angular speed of a star about the galaxy and $\kappa$ is its radial angular frequency, then we should choose $\nu_i = \nu - \nicefrac{\kappa}{2} > 0$. For near-circular orbits, $\nu_i$ will not vary much over a large region of a galaxy \citep{BinneyTremaine1987}. 

We now introduce a weak, bar-like potential, rotating with pattern speed $\nu_p \approx \nu_i$, and consider its interaction with an orbit. In the frame co-rotating with $\nu_p$, the star's orbit is an almost closed oval which rotates at a slow rate $\nu_i - \nu_p << \nu$. There is no time for the weak perturbing potential to affect the star's fast motion around the oval, so 
the orbit has an adiabatic invariant, $1/2\pi \oint \bm{p} \cdot d\bm{q} = 2J_f \sim {\rm const}$, where $\bm{q}$ is a vector of polar coordinates ($R, \phi$), $\bm{p}$ is the polar conjugate momentum, and the integral is taken over one closed, bi-symmetric orbit.  However, the potential will exert a persistent weak torque on the oval as a whole because they move slowly with respect to one another. Hence the oval will change to another oval with the same $J_f$ but different angular momentum $j$. 

If the orbit is ahead of the bar in its rotation, its angular momentum will decrease due to the gravitational torque from the bar.
Normally, $\nu_i$ will increase in response such that the orbit is repelled by the bar. In other words, the bar repels the orbit because $\nicefrac{\partial{\nu_i}}{\partial j} |_{J_f}$ (the Lynden-Bell derivative;  \citealt{Poly2004}) is negative.
In the abnormal case in which the Lynden-Bell derivative is positive however, the orbit will  oscillate about the bar-like potential. In such cases, the orbit adds to the strength of the potential which will then be able to capture more and more orbits. 

To discover what regions of a galaxy lead to the barring of near-resonant orbits, Lynden-Bell calculated $\nu_i (J_f, j)$ for an isochrone galactic potential  which permits analytic expressions for the angular frequencies $\kappa$ and $\nu$. He showed that an abnormal region is associated with central regions in this model where circular velocity rises with radius.  
\citet{Poly2020} recently expanded upon Lynden-Bell's work by mapping the equilibria of orbits as a function of $\nu_i$, the Lynden-Bell derivative, and the orbit's responsiveness to the bar potential.

We now extend this argument to a near-Keplerian system, where the gravitational potential is dominated by a central mass. A typical orbit is an almost-closed ellipse. As in \citet{lyndenbell1979} we focus on the idealized planar problem, though we note that our simulations in the next sections are three-dimensional.
The orientation of the ellipse in the orbital plane is given by the longitude of pericenter, $\varpi$, and its rate of change $\dot{\varpi}=d\varpi/dt=\nu - \kappa$ indicates its precession rate. 
If we view the orbit from rotating axes, we may choose the rotation speed such that the angle between apocenters is zero, $\nu_i = \dot{\varpi} = 0$. The orbit will then be a closed ellipse with the central body occupying one focus. 

Following Lynden-Bell's argument, we now introduce a weak, {\it lopsided} potential rotating with pattern speed $\nu_p \approx \dot{\varpi}$ and consider its gravitational influence on an orbit. 
Here the precession rate of the orbit is by definition much less than the orbital period even in an inertial frame.
In this case, $\dot{\varpi} \ll \nu$ and $\nu_p \ll \nu$, thus $\dot{\varpi} - \nu_p \ll \nu$.
Therefore, the secular average over mean anomaly is equivalent to Lynden-Bell’s average over fast orbital motion, and $J_f \rightarrow I$ where $I = \sqrt{G M a}$, $M$ is the central mass and $a$ is the semi-major axis (see \citealt{merritt2013, fouvry+2021}). 

The lopsided potential exerts a persistent torque on the orbit, changing the orbit's angular momentum at fixed semi-major axis.
The specific angular momentum of a Kepler orbit is given by $j = \sqrt{G M a (1 -e^2)}$, where $e$ is the magnitude of the orbital eccentricity. 
At fixed semi-major axis, angular momentum is a monotonically decreasing function of eccentricity.
In Kepler elements, the Lynden-Bell derivative $\left(\nicefrac{\partial{\nu_i}}{\partial j} |_{J_f}\right)$ is $\propto - \, \nicefrac{\partial{\dot{\varpi}}}{\partial e} |_{a}$.

Lynden-Bell's `abnormal region' specifically refers to prograde precession with magnitude increasing with increasing angular momentum. In Kepler elements, this corresponds to a region where precession is prograde with magnitude decreasing with increasing eccentricity. 
The interpretation of `normal' and `abnormal' regions changes with context, e.g. the abnormal region described above is actually typical in lopsided eccentric disks \citep{Madigan2018}.
Therefore, we will refer to regions where apsidal clustering is supported as {\it clustering regions}, and regions where clustering is not supported as {\it anti-clustering regions}.

We note that it is also possible to have a clustering region with retrograde precessing orbits:
if precession is retrograde and the magnitude of the precession rate increases with increasing eccentricity then orbits will be attracted to a perturbing potential.
Orbits can be trapped in modes in near-Keplerian systems provided that $\nicefrac{\partial{\dot{\varpi}}}{\partial e} |_{a} < 0$ regardless of the sign of $\varpi$; 
we define the clustering region to be any region in the disk where $\nicefrac{\partial{\dot{\varpi}}}{\partial e} |_{a} < 0$.
\section{$N$-Body Simulations}
\label{sec:methods}

Our $N$-body simulations use the open-source framework REBOUND with the IAS15 adaptive timestep integrator \citep{Rein2012, Rein2015}\footnote{The fixed timestep WHFast integrator, while faster than IAS15, doesn't conserve energy and angular momentum well in this high eccentricity problem \citep[see also][]{Rauch1999}. The performance of the MERCURIUS integrator is similar to IAS15's due to frequent close encounters between particles.}. 
Additionally, we use REBOUNDx \citep{Tamayo2019} to add a zonal harmonic, $J_2$, to the central body to emulate the orbit-averaged effects of the giant planets. 
All particles in our simulations are massive and fully-interacting.
In this paper, the Kepler elements semi-major axis ($a$), eccentricity ($e$), inclination ($i$), argument of pericenter ($\omega$), longitude of the ascending node ($\Omega$), and mean anomaly ($\mathcal{M}$) are used to describe the orbits. 

\begin{deluxetable}{ccc}
\tablecaption{Model names and parameters.}
\tablehead{
\colhead{Model ID} & \colhead{$J_2$} & \colhead{$N$} 
}
\startdata
N400 & No & 400\\
N800 & No & 800\\
J2N400 & Yes & 400\\
J2N800 & Yes & 800
\enddata
\end{deluxetable}

The total disk mass used in the simulations is $M_d = 10^{-3}\,M_\odot$ and the number of particles is 400 or 800 (see Table 1).
This unrealistically large disk mass is chosen to accelerate secular dynamics (see Equation~\ref{eq:tsec}) within the disk reducing the number of orbits we need to simulate. In addition, the low $N$ is required to reduce the simulation walltime per orbit.
The orbital distribution of our disks are initialized to approximately model a primordial scattered disk in the outer solar system \citep{Duncan1987}.
The model is axisymmetric with an order of magnitude spread in semi-major axis, $a_0$, the values of which are drawn from a 1D log-uniform distribution between $[10^2,10^3]$~AU (this is equivalent to a surface density distribution of $a^{-2}$)\footnote{We have simulated other 1D semi-major axis distributions, for example $a^{-0.7}$ \citep{Napier2021}  and $a^{-2.5}$ \citet{Duncan1987}. The instability proceeds similarly but its timescale decreases with increasing distribution steepness.}. 
All bodies have the same initial pericenter distance, $p_0= 30$~AU. Inclination $i_0$ is drawn from a Rayleigh distribution with a mean of $5^\circ$, and $\omega$, $\Omega$, and $\mathcal{M}$ are chosen uniformly from 0 to $2\pi$ radians\footnote{Disks with larger initial inclinations also undergo the instability provided that mean $i_0$ is less than $\sim20^\circ$.}. 

We add a $J_2$ potential to the central body in half of our simulations (see Table 1), and pick the $J_2$ moment to lie in the ``transition region'' where the $J_2$ potential alters the inclination instability without suppressing it \citep{Zderic2020b}. 
Our chosen disk mass and number of particles, choices forced by numerical limitations, are unrealistic. The instability timescale and the max $J_2$ that the disk can resist (that is, still undergo the instability) both depend on these key parameters. The low $N$ in our simulations leads to artificially strong self-stirring that weakens the secular torques that cause the inclination instability and increases differential precession by excessively spreading out the disk \citep{Madigan2018b}. For the same total mass, a disk with more particles will be able to resist a larger $J_2$ (that is, still undergo the instability).
We determined how the inclination instability timescale scales with $M_d$ and $N$ in \citet{Madigan2018b}. Then in \citet{Zderic2020b}, we used that timescale scaling along with simulations of these disks with added $J_2$ to find that a $\sim20\,M_\oplus$ primordial scattered disk could resist the $J_2$ of the giant planets. We found that this realistic system would be in the transition region. The J2N400 and J2N800 models in this paper are in the transition region too. Therefore, these models, which have unrealistic $J_2$, $M_d$, and $N$, are dynamically similar to a 20 Earth mass primordial scattered disk, at least with regards to $J_2$.

For the sake of reproducibility, the $J_2R^2$ used in these simulations is $0.3\,{\rm AU}^2$. We use the same $J_2 R^2$ value for the J2N400 and J2N800 even though these simulations have different critical $J_2$ because this $J_2 R^2$ is sufficient to put both models in the transition region. The $J_2 R^2$ for the solar system is 0.06 ${\rm AU}^2$ and using solely a secular scaling ($10^{-3}/20M_\oplus \approx 16$) $J_2 R^2$ would be 0.96 AU$^2$ for a $10^{-3} M_\odot$ mass disk. This $J_2 R^2$ is about 3 times larger than the actual $J_2 R^2$ used
in our simulations. A $N \rightarrow \infty$ disk can resist about 3 times more $J_2 R^2$ than a $N = 400$ disk.

Simulation times are given in units of the secular timescale: 
\begin{equation}
    \label{eq:tsec}
    t_{\rm sec} = \frac{1}{2\pi}\frac{M_\odot}{M_{\rm d}}P
\end{equation}
where $P$ is the orbital period at the innermost part of the disk.
For $M_d = 10^{-3}\,M_{\oplus}$, $t_{\rm sec} \approx 160\,P \approx 0.16\,{\rm Myr}$, where $P(a = 100 \,{\rm AU}) = 10^3 \,{\rm yr}$.
We give timescales for a more realistic 20 Earth mass primordial scattered disk with in Section~\ref{sec:conclusions}.

\section{Results}
\label{sec:results}

\begin{figure*}[!htb]
	\centering
	\includegraphics[width=\textwidth]{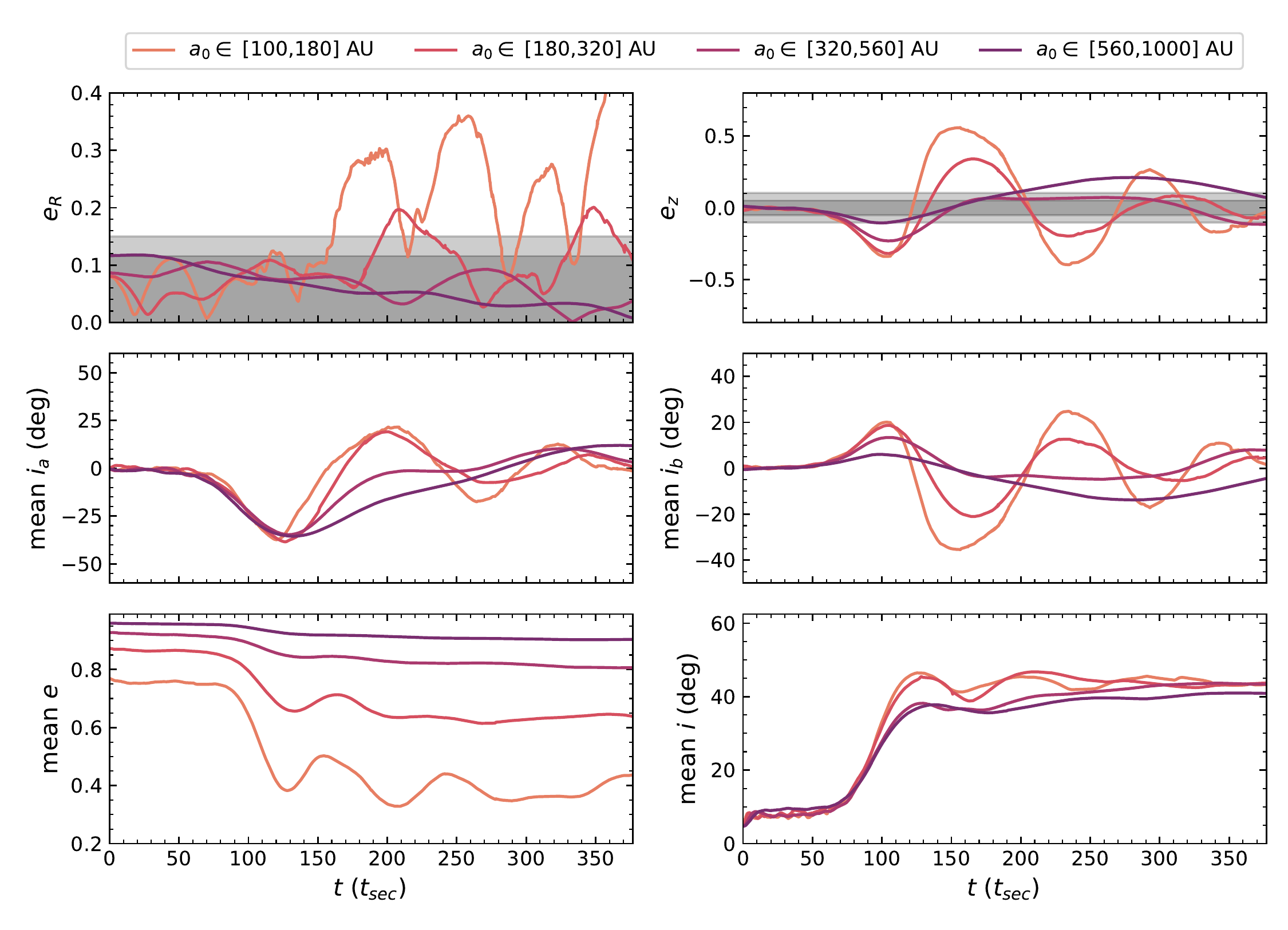}
	\caption{Model N400: Apsidal clustering occurring in the inner edge of a scattered disk model without added $J_2$ after the inclination instability has saturated. The plotted quantities are binned by initial semi-major axis. Two noise floors are shown for $e_R$ and $e_z$ (see Section~\ref{sec:results} for noise floor definition). The inclination instability is shown by exponential growth in $i$, $i_a$, and $i_b$ and a corresponding decrease in $e$ to conserve total angular momentum of the disk. The instability saturates at $t\sim125\,t_{\rm sec}$. About $25\,t_{\rm sec}$ later, $e_R$ for the inner bin ($a_0 \in [100,180]$ AU) increases above the noise floor indicating in-plane apsidal clustering. About $50\,t_{\rm sec}$ later, slight in-plane apsidal clustering appears in the next bin ($a_0 \in [180,320]$ AU).}
	\label{fig:sd-N400}
\end{figure*}

\begin{figure*}[!htb]
	\centering
	\includegraphics[width=\textwidth]{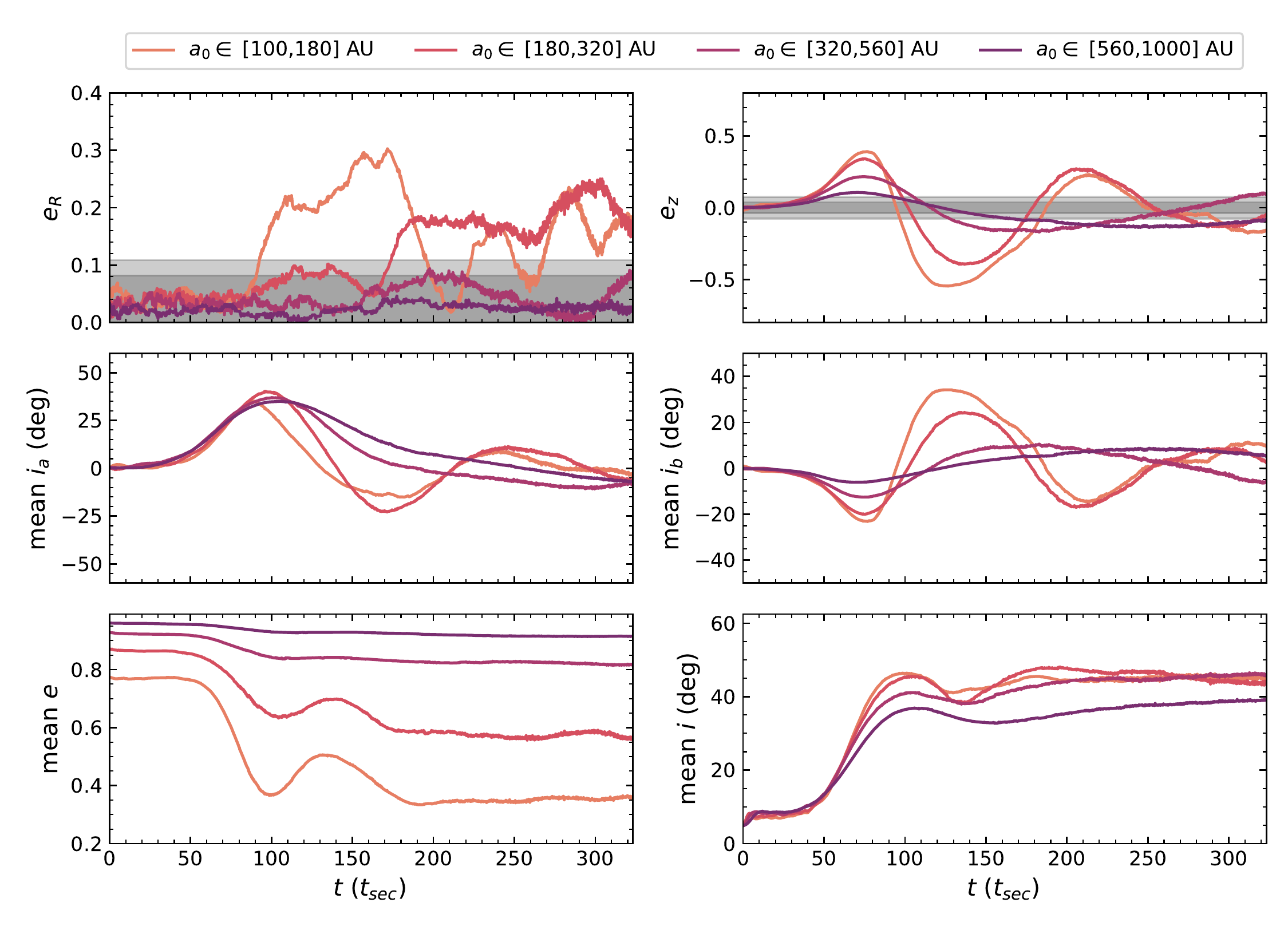}
	\caption{Model N800: Apsidal clustering occurring in the inner edge of an 800 particle scattered disk model without added $J_2$ after the inclination instability has saturated. Same panels as in Figure~\ref{fig:sd-N400}. The plotted quantities are binned by initial semi-major axis, and $e_R$ and $e_z$ noise floors are shown. Compared to N400, the instability saturates at an earlier time, $t\sim100\,t_{\rm sec}$, and, in-plane apsidal clustering (shown by $e_R$) in the inner $a_0$ bin begins immediately after the instability saturates. Like N400, in-plane apsidal clustering propagates into the next $a_0$ bin ($a_0 \in [180,320]$ AU) about $75\,t_{\rm sec}$ later.
    Apsidal clustering in the $a_0 \in [180,320]$ AU bin is stronger and more consistent in N800 than in N400.}
	\label{fig:sd-N800}
\end{figure*}

\begin{figure*}[!htb]
	\centering
	\includegraphics[width=\textwidth]{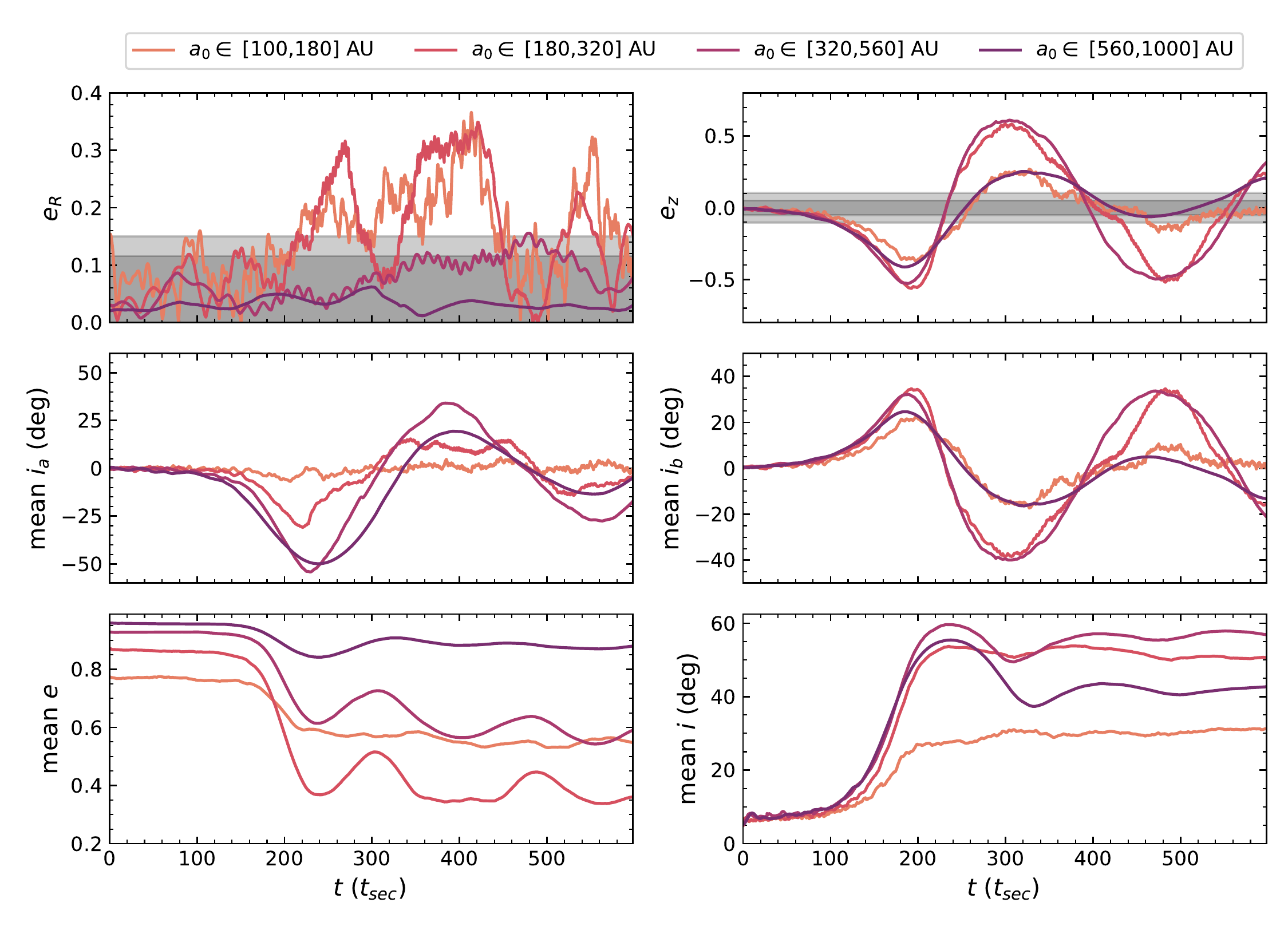}
	\caption{Model J2N400: Apsidal clustering occurring in the inner edge of a scattered disk model with added $J_2$ after the inclination instability has saturated. Same panels as in Figure~\ref{fig:sd-N400}. Compared to N400 and N800, the inclination instability is delayed by the added $J_2$, saturating at $\sim200\,t_{\rm sec}$, and the innermost part of the disk barely undergoes the instability. Like in N400 and N800, a lopsided mode develops on the inner edge of the disk shortly after the inclination instability has saturated. Unlike N400 and N800, apsidal clustering appears in both inner disk bins $a_0 < 320$~AU simultaneously. In addition, apsidal clustering in the $a_0 \in [180,320]$ bin is as strong as in the $a_0 \in [100,180]$ bin.}
	\label{fig:sd-J2N400}
\end{figure*}

\begin{figure*}[!htb]
	\centering
	\includegraphics[width=\textwidth]{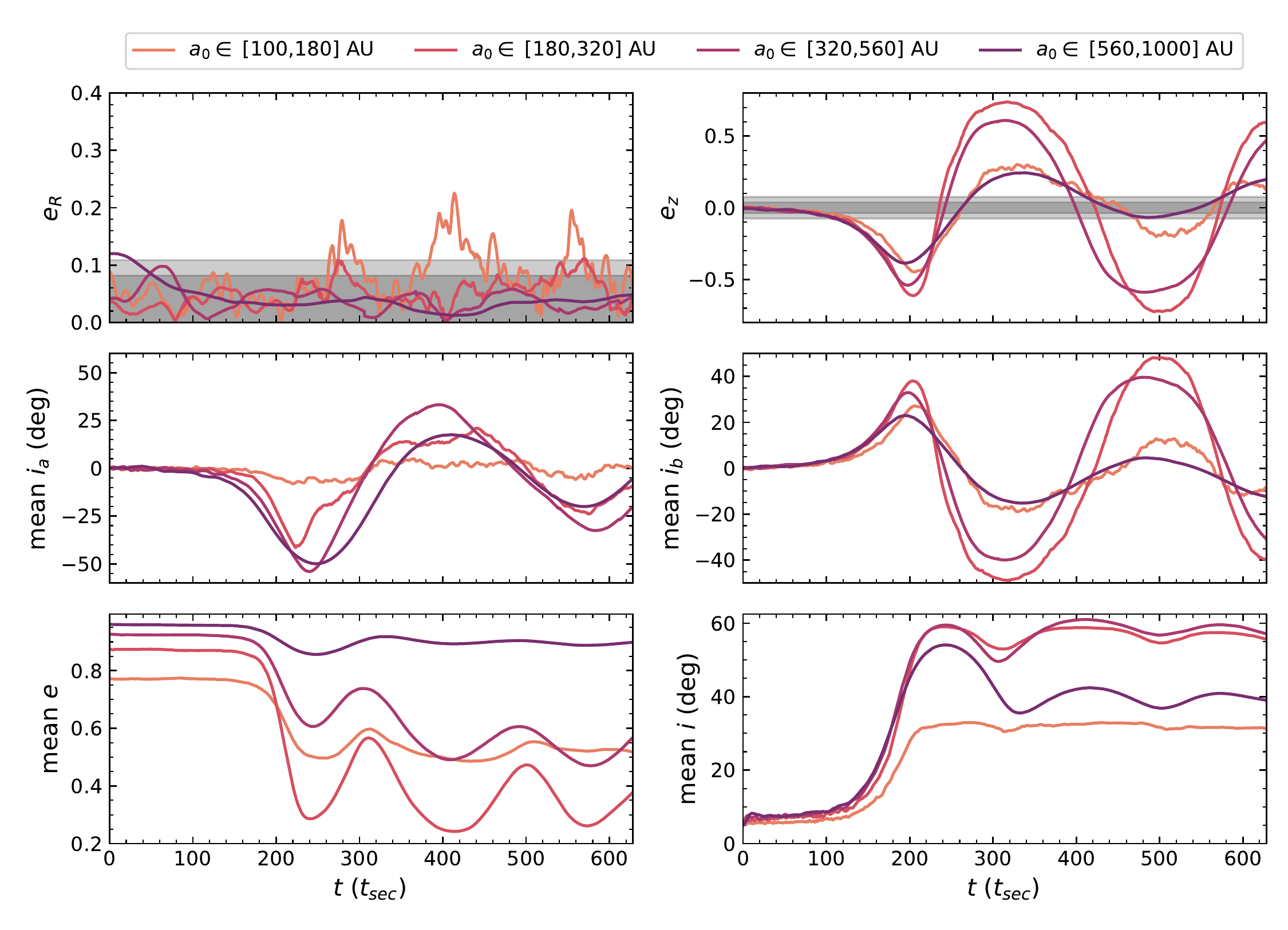}
	\caption{Model J2N800: Apsidal clustering occurring in the inner edge of an 800 particle scattered disk model with added $J_2$ after the inclination instability has saturated. Same panels as in Figure~\ref{fig:sd-N400}. Compared to J2N400, apsidal clustering is weakened in this simulation. Statistically significant clustering only occurs in the $a_0 \in [100,180]$ AU bin and this clustering is weaker than in the J2N400. This is different than the simulations without added $J_2$ where we saw similar to slightly more in-plane apsidal clustering as we increased the in particle number.}
	\label{fig:sd-J2N800}
\end{figure*}

\begin{figure*}[!htb]
	\centering
	\includegraphics{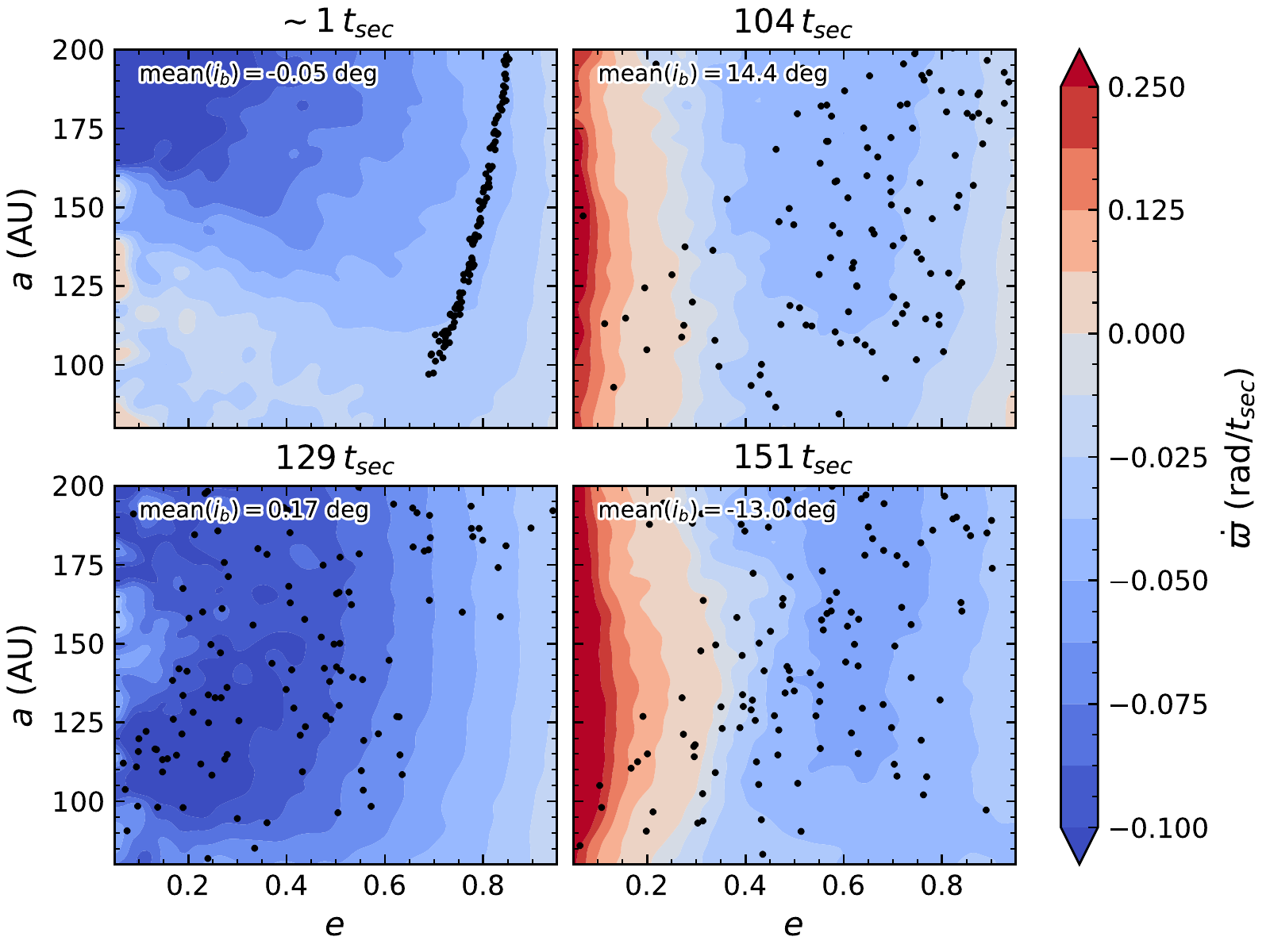}
	\caption{Model N400: Contours of the time derivative of the longitude of pericenter, $\dot{\varpi}$, for the inner edge ($a \in [80,200]\,{\rm AU}$) of the scattered disk simulation without $J_2$ from Figure~\ref{fig:sd-N400} at four times.
	The $e$ and $a$ of the disk particles are shown with black points and the mean $i_b$ of the disk at the time is shown in the top left.
	There is a clustering region at $t=104\,t_{\rm sec}$ and $t = 151\,t_{\rm sec}$ for $e\in[0.25,0.60]$. This clustering region is associated with the bowl-shaped orbital configuration created by the inclination instability. Note that the region of prograde precession, $e\in[0.05,0.25]$, also supports apsidal clustering as described in Section~\ref{sec:LB}. Apsidal clustering doesn't appear until after $151\,t_{\rm sec}$. At $104\,t_{\rm sec}$ the region $e\in[0.05,0.60]$ is only occupied by $\sim20$ particles, by $151\,t_{\rm sec}$ this has increased to $\sim65$ particles. There aren't enough particles in $e\in[0.05,0.60]$ at $104\,t_{\rm sec}$ for apsidal clustering to begin earlier.}
	\label{fig:scatdisk-contour}
\end{figure*}

We measure apsidal clustering using the mean, normed eccentricity vector,
\begin{equation}
    \bm{\mu}_{\hat{\bm{e}}} = \sum_{i=1}^N \frac{\hat{\bm{e}}_i}{N}, 
\end{equation}
where 
\begin{equation}
\bm{e}_i = \frac{(\bm{v}_i \times \bm{j}_i)}{ GM_\odot} - \hat{\bm{r}}_i 
\end{equation}
is the eccentricity vector of the $i$th orbit, and $\bm{r}_i$, $\bm{v}_i$, and $\bm{j}_i$ are the position, velocity, and specific angular momentum of the $i$th particle. 
The eccentricity vector points from the apocenter to the pericenter of the orbit.
We use the cylindrical coordinates of $\bm{\mu}_{\hat{\bm{e}}}$ to look for apsidal clustering,
\begin{subequations}
\begin{align}
  e_R &= \sqrt{\mu_{\hat{\bm{e}},x}^2 + \mu_{\hat{\bm{e}},y}^2}, \\
  e_\theta &= \arctan{\left[ \frac{\mu_{\hat{\bm{e}},y}}{\mu_{\hat{\bm{e}},x}} \right]} , \\
  e_z &= \mu_{\hat{\bm{e}},z}.
\end{align}
\label{eq:eReTez}
\end{subequations}
The radial component, $e_R$, quantifies in-plane apsidal clustering, the azimuthal component, $e_\theta$, is used to calculate the pattern speed and direction of in-plane apsidal clustering, and the $z$ component, $e_z$, quantifies out-of-plane apsidal clustering. See Appendix~\ref{app:measure-apsidal} for a comparison to standard measures of apsidal clustering.

We calculate the noise floor for $e_R$ and $e_z$ by creating one thousand $N=400$ or $N=800$ axisymmetric disks. For each disk, we draw argument of perihelia and longitude of the ascending node from a uniform distribution and inclination from a Rayleigh distribution with mean inclination equal to the mean inclination of the post-instability disk (e.g. $\sim 50 \, {\rm deg}$). We calculate $e_R$ and $e_z$ for each disk to obtain an empirical distribution for $e_R$ and for $e_z$ with $N=1000$ samples. We calculate the noise floors from these distributions (68th and 95th percentile centered on the mean, corresponding to one and two standard deviations of the Gaussian distribution).
The noise floor is a function of $N$ with lower $N$ simulations having higher noise floors.
We show the noise floors in Figures~\ref{fig:sd-N400}, \ref{fig:sd-N800}, \ref{fig:sd-J2N400} and  \ref{fig:sd-J2N800} with grey bands.
$e_R$ and $e_z$ values above the noise floor indicate statistically significant apsidal clustering. 

As first described in \citet{Madigan2016}, the spatial orientation of orbits can be quantified with the angles, \ia, \ib, and \ie representing rotations of an orbit about its semi-major (${\hat a}$) axis, semi-minor (${\hat b} \equiv \hat{j} \times \hat{a}$) axis, and angular momentum vector ($\hat{j}$), respectively, such that
\begin{subequations}
\begin{align}
  \ia &= \arctan\left[\unitvectorslope{b}\right], \\
  \ib &= \arctan\left[-\unitvectorslope{a}\right], \\
  \ie &= \arctan\left[\oldhat{a}_{\text{y}}, \oldhat{a}_{\text{x}}\right].
\end{align}
\label{eq:iaibie}
\end{subequations}
\\
The subscripts $x$, $y$, and $z$ denote an inertial Cartesian reference frame with unit vectors, $\hat{x}$, $\hat{y}$, and $\hat{z}$.
The angles \ia, \ib, and \ie are equivalent to the roll, pitch and yaw of a boat or plane, and are useful for understanding the net gravitational torque acting on an orbit. 
The inclination instability is characterized by the mean $i_a$ (roll) and $i_b$ (pitch) of all the orbits in the disk growing exponentially with opposite signs. 
We use these angles in upcoming plots to see how the inclination instability proceeds in simulations with different parameters and how that affects the subsequent growth of a lopsided mode.

\subsection{Inclination Instability}
\label{sec:inc-ins}

The axisymmetric  disks of eccentric orbits in our simulations undergo a dynamical instability called the inclination instability due to the secular gravitational torques between orbits. The instability is characterized by exponential growth in inclination and a corresponding decrease in eccentricity. The initially thin disk expands into a cone or bowl shape\footnote{For a visualization of the bowl shape, the reader can jump ahead to the second row, third panel from the left of Figure~\ref{fig:observables}.}. As the orbits’ inclinations grow, they tilt in the same way with respect to the disk plane and oscillate coherently in $i_a$ and $i_b$. We describe the physical mechanism behind the inclination instability in \citet{Madigan2018b}.

In Figures~\ref{fig:sd-N400}, \ref{fig:sd-N800}, \ref{fig:sd-J2N400}, and \ref{fig:sd-J2N800}, we show the inclination instability and its aftermath for models N400, N800, J2N400, and J2N800, respectively. 
Particles are binned by their initial semi-major axis, with the bin boundaries chosen such that the number of particles per bin is approximately equal.\footnote{We've verified that particles do not drift far from their initial semi-major axis during integration.}
Note that the figures have different $x$-axes (time) but identical $y$-axes. 

In Figures~\ref{fig:sd-N400} and \ref{fig:sd-N800} (models N400 and N800), the largest growth in inclination occurs in the two innermost semi-major axis bins. 
In Figures~\ref{fig:sd-J2N400} and \ref{fig:sd-J2N800} (models J2N400 and J2N800) however, the innermost bin ($a < 180$ AU) flattens in inclination after a shorter exponential phase. 
This difference is seen again in the eccentricity evolution; the innermost semi-major axis bin drops to the lowest eccentricity values in the simulations without added $J_2$ whereas the drop is suppressed in the simulations with added $J_2$.
In addition, the inclination instability saturates at later times in the J2N400 and J2N800 models than in N400 and N800 models ($\sim200\,t_{\rm sec}$ vs. $\sim 100\,t_{\rm sec}$).
We attribute the difference between the two models to the strong differential apsidal precession in the innermost bin induced by the gravitational influence of the giant planets. 
This effect decreases the coherence time over which inter-orbit torques can act.

The inclination instability produces \textit{out-of-plane} apsidal clustering, captured by both $e_z$ and $i_b$. 
We note that the longitude of pericenter, $\varpi = \omega + \Omega$, fails to find this clustering because it is sensitive only to in-plane clustering. 

\subsection{Apsidal Clustering in the Scattered Disk}
\label{sec:scat-disk-cluster}

In \citet{zderic2020a}, we found \textit{in-plane} apsidal clustering after the inclination instability had saturated in a simple, unrealistic orbital configuration. This ``compact configuration'' is characterized by an axisymmetric, nearly-flat disk of Keplerian orbits in which all bodies have identical eccentricities and nearly identical semi-major axes. Here, we report the same findings for our scattered disk model with and without added $J_2$.

In Figures~\ref{fig:sd-N400} and \ref{fig:sd-N800}, $e_R$ traces the development of apsidal clustering in the $x/y$-plane at the inner edge ($a_0\in[100,180]$ AU) of Models N400 and N800\textemdash a massive scattered disk without giant planets. Values of $e_R$ above the noise floor indicate statistically significant in-plane apsidal clustering.
As in \citet{zderic2020a}, apsidal clustering appears in the disk after the inclination instability has saturated.
Note that apsidal clustering only appears for bodies with $a \lesssim 320$~AU with clustering first appearing in the $a_0 \in [100,180]$~AU bin and then travelling out into the $a_0 \in [180,320]$~AU bin at later times.
Comparing the two models, apsidal clustering begins earlier, is more consistent (fewer oscillations), and is stronger in the $a_0 \in [180,320]$~AU bin in the higher $N$ model, N800, than in the N400 model.
Finally, note that the mode strength regularly oscillates below the noise floor, particularly in N400.

Figures~\ref{fig:sd-J2N400} and \ref{fig:sd-J2N800} show the development of apsidal clustering in the inner edge of our $J_2$ models, J2N400 and J2N800\textemdash a massive scattered disk with giant planets. 
We get apsidal clustering in both models (starting after $\sim200\,t_{\rm sec}$), though this clustering is weaker than it is for the models without added $J_2$. 
Apsidal clustering appears at later times in the $J_2$ models because the inclination instability is slowed by the added $J_2$, and clustering does not appear until after the instability has saturated. 
In J2N400, apsidal clustering in the $a_0\in[180,320]$~AU bin is stronger than it is in the N400 and N800 models.
This reflects a general trend of our $J_2$ results.
The $J_2$ potential disrupts the instability for the lowest $a$ bodies (compare the $a_0 \in [100, 180]$~AU bin in Figures~\ref{fig:sd-N400} and \ref{fig:sd-J2N400}), but it strengthens the instability in the outer $a_0$ bins.
Bodies with $a_0 \gtrsim 180$~AU attain higher mean $i$, lower mean $e$, and stronger apsidal clustering post-instability in the J2N400 model than in N400 and N800 models.
In J2N800, statistically significant apsidal clustering only appears in the $a_0 \in [100,180]$~AU bin, and it's weaker and shorter-lived than all the other models. 
This is unexpected\textemdash clustering was generally stronger in N800 than in N400 so we expect J2N800 to show apsidal clustering similar to or stronger than J2N400.
We have multiple simulations of the $N=400$ models, all showing apsidal clustering.
The N800 and J2N800 simulations we show here are the only ones of that particle number that were run long enough to show apsidal clustering.
In the $N=400$ simulations, we find that the strength of clustering varies from simulation to simulation (being as weak as $e_R \approx 0.20$ at peak).
Thus, it is possible that the weak apsidal clustering seen in J2N800 is just a peculiarity of that simulation's specific initial conditions.

In all models, in-plane apsidal clustering appears in the inner semi-major axis bin(s) after the instability has saturated.
The occurrence of apsidal clustering shortly after the inclination instability in both models suggests that this instability is responsible for the in-plane apsidal clustering. 

\begin{figure*}[!htb]
	\centering
	\includegraphics{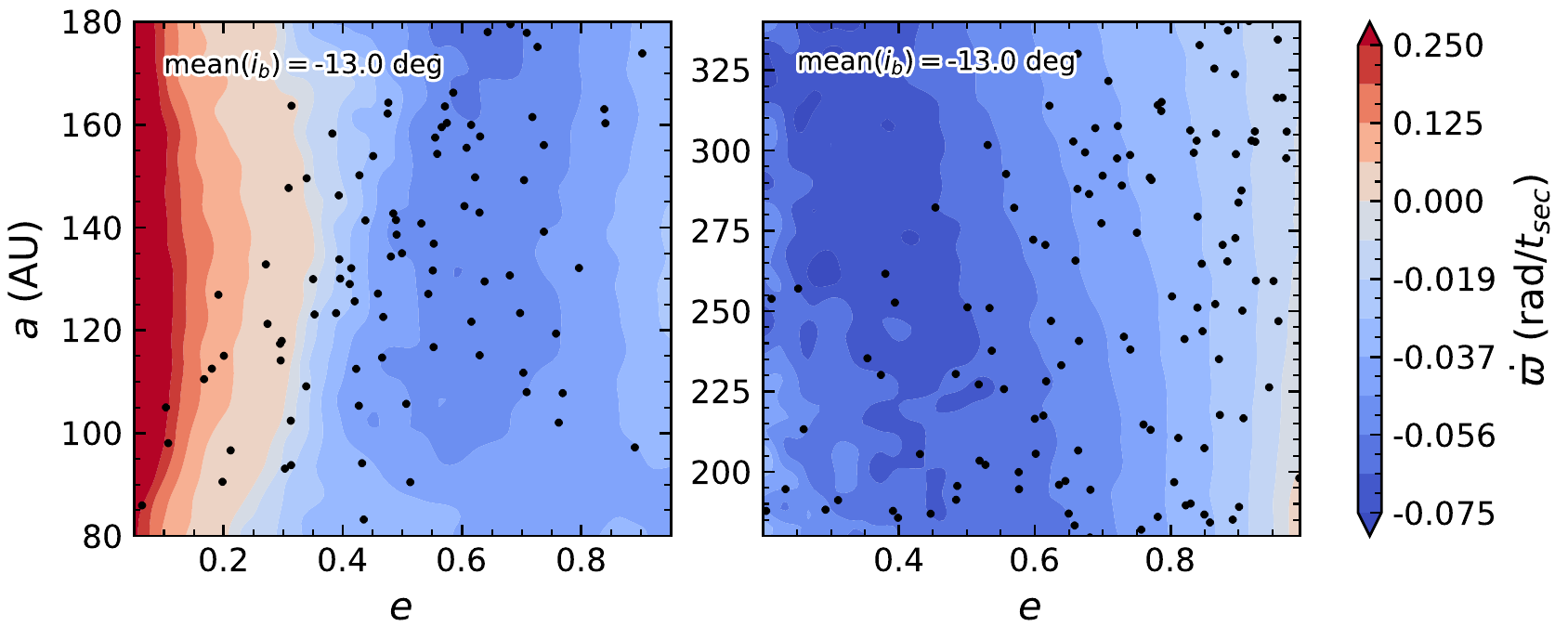}
	\caption{Model N400: $\dot{\varpi}$ contour plots for the innermost two $a_0$ bins at $t=151\,t_{\rm sec}$. The left panel is the same as the bottom right panel in Figure~\ref{fig:scatdisk-contour}, and it shows a large clustering region ($e \lesssim 0.6$). The right panel shows a small, underpopulated clustering region at the lowest  $a$ (note that the $e$ axes are different in the two panels). This explains why we only see strong apsidal clustering for $a \lesssim 180\,{\rm AU}$. The contour plots differ at their boundary, 180 AU. This is due to the different mean test particle $\omega$ in these two bins and it demonstrates the importance of $\omega$ in forming the clustering region.}
	\label{fig:scatdisk-contour-bin}
\end{figure*}

\begin{figure*}[!htb]
	\centering
	\includegraphics[width=\textwidth]{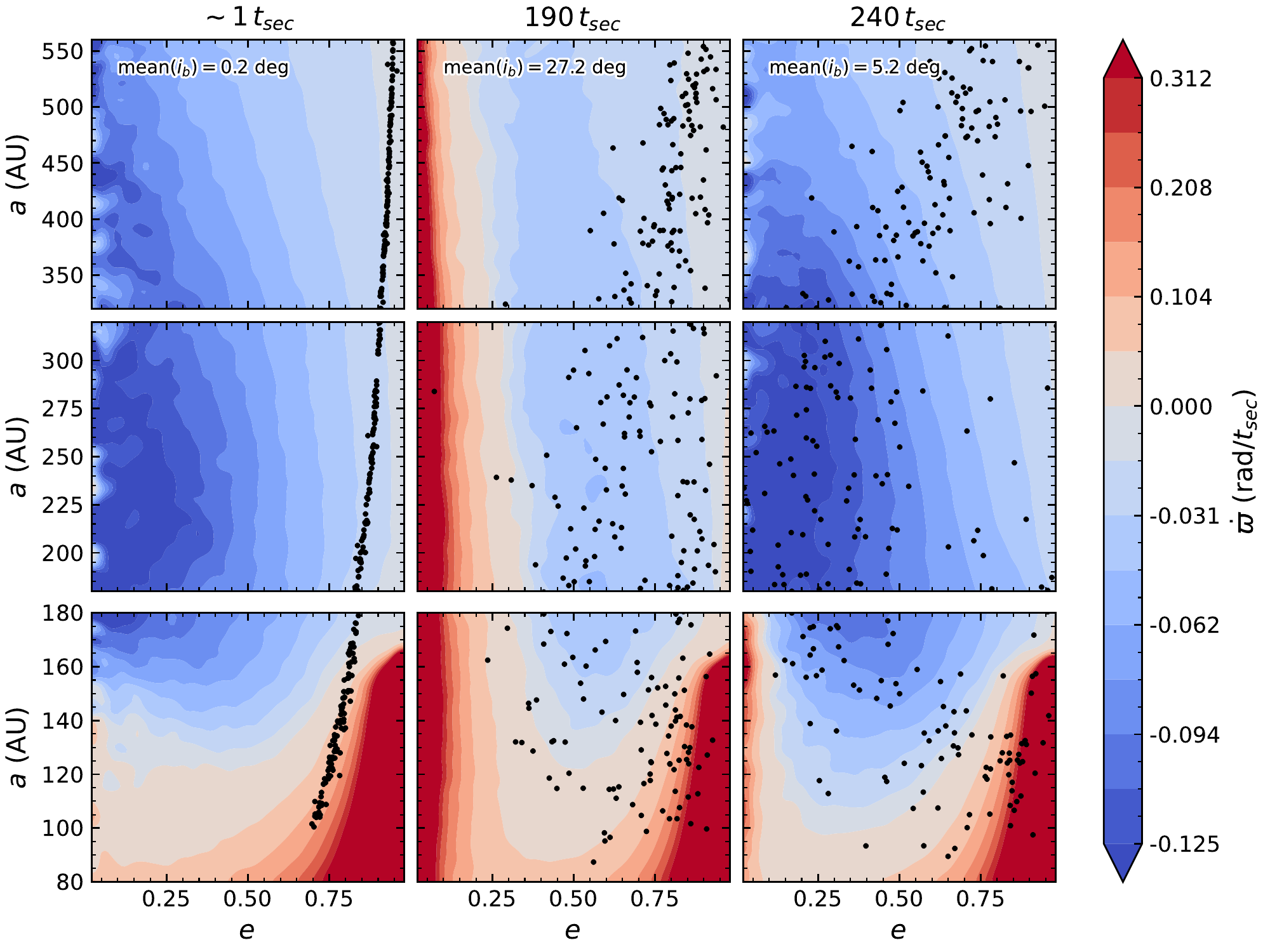}
	\caption{Model J2N400: The emergence of a clustering region post-instability in a scattered disk model including $J_2$ (the same simulation as shown in Figure~\ref{fig:sd-J2N400}). We show contours of $\dot{\varpi}$ from three different $a$ bins ($a_0 \in [100, 180]$ AU, $a_0 \in [180, 320]$ AU and $a_0 \in [320,560]$ AU) at three different times. Mean $i_b$ for the disk is shown in the top left of the top panel. Peak mean $i_b$ is attained at $190\,t_{\rm sec}$, and a corresponding, populated clustering region is shown for $e \in [0.05, 0.55]$ and $a \in [125,300]\,{\rm AU}$. The deep trough of prograde precession at high $e$ and low $a$ in the bottom row of panels is due to the $J_2$ potential.}
	\label{fig:scatdisk-contour-J2}
\end{figure*}

\subsection{Emergence of the clustering region}
\label{sec:clustering-region-emerges}

Apsidal clustering in our simulations occurs after a Lynden-Bell clustering region appears. Here we show that the clustering region appears during the saturation of the inclination instability and is associated with the unique, `bowl'-shape of the mass distribution post-instability.

We simulate test particles in the frozen gravitational potential of the disk to find $\dot{\varpi}$ as a function of $a$ and $e$.
The disk particles from a fully interacting simulation are frozen onto their orbits at a specific time and a test particle is integrated in this background disk potential.
We use these test particles to make contour plots of $\dot{\varpi}$ in $e$-$a$ space at specific instances to find clustering regions, if any exist. 
In these test particle simulations we have a test particle, a background disk, and a central body (with/without added $J_2$). 
The background disk particles are given the REBOUND/MERCURY `small particle' designation meaning they do not interact with each other; they only interact with the test particle and the central body (they do actually interact with each other indirectly through the central body as mentioned in \citet{Peng2020}; this effect is small).
Paradoxically, we must give the test particle mass in order for it to interact with the background disk. 
The mass of the test particle is set to be so small that it negligibly affects the background disk bodies.
The test particle simulations are integrated for 10 orbits, and the test particles are initialized with $a$ and $e$ drawn from a grid ($96 \times 96$), $\omega$ and $i$ calculated from the mean values of the local disk (same $a$ bin), and an $\Omega$ of 0, $\nicefrac{\pi}{4}$, $\nicefrac{\pi}{2}$, or $\nicefrac{3\pi}{4}$. 
The median $\dot{\varpi}$ is calculated from each set of four test particle simulations. 
This is the method used to create the contour plots shown in Figures~\ref{fig:scatdisk-contour}, \ref{fig:scatdisk-contour-bin}, and \ref{fig:scatdisk-contour-J2}.
We have checked the accuracy of our test particle simulations with an alternative method which calculates the instantaneous precession rate of the test orbit directly from the torques and forces it experiences; the results are are consistent, as shown in Appendix~\ref{app:finding-clustering}.
We only show $\dot{\varpi}$ for the $N=400$ simulations as the $N=800$ results are the same.

In Keplerian elements, a clustering region (see Section~\ref{sec:LB}) is defined by $\nicefrac{\partial{\dot{\varpi}}}{\partial{e}} |_{a} < 0$.
Apsidal precession within our disks is initially retrograde ($\dot{\varpi} < 0$) and with {\it magnitude} decreasing with increasing eccentricity ($\nicefrac{\partial{\dot{\varpi}}}{\partial{e}} |_{a} > 0$).
When clustering regions appear within our disks, we see regions where the precession is retrograde with increasing magnitude and regions where precession is prograde with decreasing magnitude. 
In the contour plots, clustering regions are regions where, at fixed semi-major axis ($a$), the contours go from warmer to cooler colors with increasing eccentricity ($e$).

In Figure~\ref{fig:scatdisk-contour}, we show the development of a clustering region in the inner edge of N400, the scattered disk model without added $J_2$. This figure shows the time derivative of the longitude of pericenter, $\dot{\varpi}$, in the scattered disk as a function of semi-major axis $a$ and orbital eccentricity $e$ for the inner edge of the disk. 
We show mean $i_b$ for the disk in the top left of the panels.
Initially, all bodies in the scattered disk are on the line $a\,(1-e) = 30\,{\rm AU}$ (top left). 
Later, the inclination instability reduces the disk orbits' eccentricity, $e$, at roughly fixed semi-major axis, $a$ (top right), and causes the disk to buckle in to a bowl-shape.
Notably, the $\dot{\varpi}$ contours have changed to admit a clustering region $\left(\nicefrac{\partial{\dot{\varpi}}}{\partial{e}} |_{a} < 0\right)$ covering $e \in [0.25, 0.60]$ and $a \in [80, 200]\,{\rm AU}$. 
This retrograde clustering region smoothly blends into a prograde region ($e \in [0.05, 0.25]$) which also facilitates apsidal clustering. 
Thus the whole region $e \in [0.05, 0.60]$ supports apsidal clustering.
Immediately after the instability saturates, the apsidal clustering region is lightly populated. 
The eccentricity continues to drop after the inclination instability leaves the linear regime (bottom left). 
However, the clustering region has disappeared.
This is because the disk has precessed out of the bowl-shape (mean $i_b \sim 0^\circ$).
Finally, the orbits at the inner edge have precessed through the ecliptic and inverted the bowl-shaped mass distribution (mean $i_b > 0$) (bottom right).
Again, the clustering region appears, but now it is sufficiently populated for in-plane apsidal clustering to take hold.

Two things are apparent from this sequence.
First, the clustering region coincides with the bowl-shaped orbital distribution (large mean $i_b$).
Second, in-plane apsidal clustering only appears once the clustering region is sufficiently populated.

Once apsidal clustering has been established and the lopsided mode has grown, it is no longer reliant on the clustering region produced by the bowl-shape to exist. 
The bowl-shape is not actively maintained after the inclination instability saturates. 
In Figure~\ref{fig:sd-N400}, differential precession slowly erodes mean $i_b$ in the disk, and will eventually erase the bowl-shape altogether.
However, $e_R$ appears unaffected by this, and apsidal clustering actually reaches peak strength by the end of the simulation even though the mean $i_b$ has dropped quite low.
The bowl-shape seeds the lopsided mode, but, once seeded, the mode is self-sustaining even though the strength of the mode oscillates. 

The clustering region appears towards the inner edge of the disk in N400.
In Figure~\ref{fig:scatdisk-contour-bin}, we show the inner two semi-major axis bins of the disk ($a \in [100, 320]\,{\rm AU}$) at $151\,t_{\rm sec}$.
The clustering region extends to $a>200$~AU, but it is largest at lower $a$ and it's unpopulated for $a \gtrsim 250$~AU.
This explains why apsidal clustering is primarily only seen in the inner two bins of these simulations, and why apsidal clustering is slightly weaker for $a_0 > 180$~AU.
The precession rates are discontinuous at 180 AU (top of the left panel and bottom of the right panel), and the two panels have different $x$ axes, exacerbating the apparent discontinuity. 
The discontinuity is due to the test particles in these two panels having different bin-averaged $\omega_0$ and $i_0$.

The general features found in the N400 model are repeated in the J2N400 model: the clustering region appears around peak mean $i_b$ (in the `bowl'-shape), the semi-major axis location of the clustering region traces apsidal clustering, and the clustering region precedes apsidal clustering.
In Figure~\ref{fig:scatdisk-contour-J2}, we show contours of $\dot{\varpi}$ across the disk at three distinct times for model J2N400, the same simulation depicted in Figure~\ref{fig:sd-J2N400}.
A deep trough of prograde $\varpi$ precession from the added $J_2$ is seen at large eccentricity and small semi-major axis. 
Bodies near this trough do not undergo the inclination instability and their eccentricities are stable.  
The clustering region still forms (middle column), but at slightly larger semi-major axis because the lower semi-major axis portion of the disk is too disrupted by the $J_2$ potential.
This is reflected in Figure~\ref{fig:sd-J2N400} by apsidal clustering in $a_0 \in [100,180]\,{\rm AU}$ and $a_0 \in [180,320]\,{\rm AU}$.
The mean $i_a$ of the inner edge of the disk is $\sim0$ (due to the added $J_2$ precession), but mean $i_b > 0$. 

\begin{figure*}[!htb]
	\centering
	\includegraphics[scale=1.0]{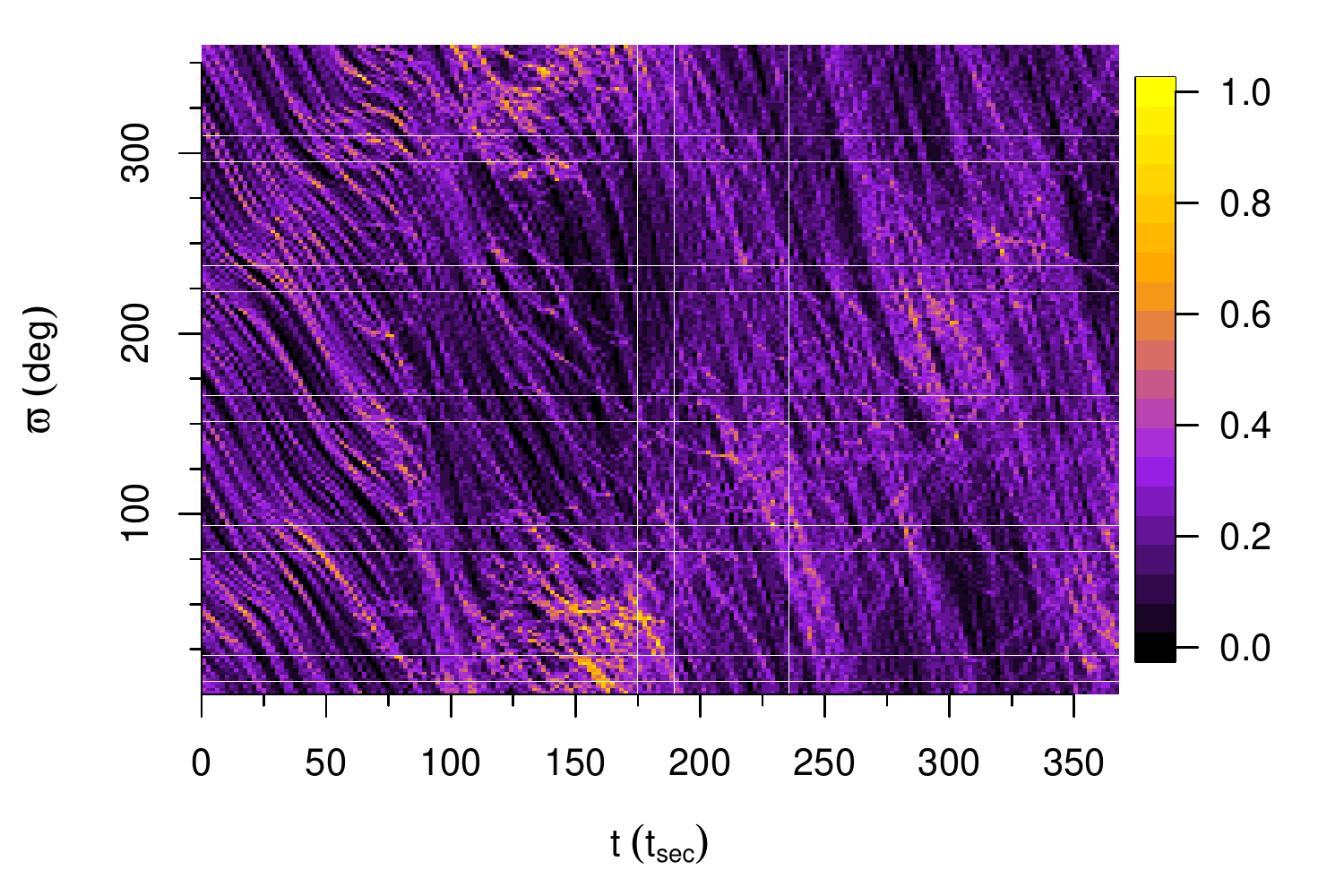}
	\caption{Model N800: Clustering in $\varpi$ as it evolves over time in the inner $a_0$ bin ($a_0 \in [100, 180]$ AU). This figure shows a 2d-histogram of $\varpi$ as a function of time for all particles in this bin, and the colors represent the density of points (normalized to the densest area).  We identify the mode as the highly dense region in this figure that generally precesses prograde, opposite to the individual orbits that precess retrograde. Retrograde precessing orbits can be captured in the prograde precessing mode, and orbits actually caught in the mode librate within the mode. The pattern speed of the mode is $\sim1.5^\circ\,t_{\rm sec}^{-1}$ which is of the same order of magnitude as the precession rate of the orbits in the disk.}
	\label{fig:mode-dynamics}
\end{figure*}

In our simulations, the formation of the clustering region coincides with peak mean $i_b$, signifying that the unique out-of-plane orbital distribution resulting from the inclination instability is responsible for the in-plane apsidal clustering. The post-instability bowl-shaped potential creates a clustering region at low eccentricity, and this region is simultaneously populated by the circularizing effect of the instability.

    A particle's precession rate depends on the forces it experiences throughout its orbit. From equation~\eqref{eq:ederiv} the orbital precession rate is
        \begin{equation}
            \dot{\varpi} \approx \left<f_r(r)\right> \frac{\sqrt{1-e^2}}{e} \sqrt{\frac{a}{G M}},
            \label{eq:precApprox}
        \end{equation}    
        where $\left<f_r(r)\right>$ is the orbit-averaged specific, radial force. This is negative if the force is radially inwards, such that precession will be retrograde. For retrograde precession, the magnitude of $\dot{\varpi}$ must increase with eccentricity in a clustering region. In contrast, the second term on the right-hand-side of equation~\eqref{eq:precApprox} decreases monotonically with eccentricity. Therefore, the orbit-averaged force must increase with eccentricity in a retrograde clustering region. This is somewhat unusual considering that typically in a Keplerian potential (i) precession is dominated by the forces near apocenter and (ii) these forces will be smaller at larger apocenters. However, this is not the case in the post-instability bowl-shaped orbital distribution. 
        
        In Appendix~\ref{app:finding-clustering}, we measure the precession rates of orbits in the clustering region, by calculating the forces and torques at many points along the orbit. We find that changes in the precession rate (with eccentricity) can be dominated by points near pericenter. Additionally, forces can increase with apocenter. This behavior allows a clustering region to appear. 
        
        For prograde precession, clustering will occur if the magnitude of $\dot{\varpi}$ decreases with eccentricity, which is typical for a Keplerian potential. In fact this will occur for any external force that decreases with radius (see equation~\ref{eq:precApprox}).

\subsection{Mode direction} 
\label{subsec:mode-dir}

In galaxies, the slowness condition, $\nu_i - \nu_p \ll \nu$, will only be satisfied if the mode and the orbits precess in the same direction because $\nu_i \lesssim \nu$ (galactic orbits are generally rosettes in the inertial frame).
If ${\rm sgn}(\nu_i) = -{\rm sgn}(\nu_p)$ then $\nu_i - \nu_p \sim \nu_i$ and the orbits are not nearly closed in the rotating frame of the bar perturbation.
However, in near-Keplerian systems, both $\dot{\varpi}_i$ and $\nu_p$ are much less than $\nu$ and $\dot{\varpi}_i - \nu_p \ll \nu$ is true even if ${\rm sgn}(\dot{\varpi}_i) = -{\rm sgn}(\nu_p)$.
If the mode and the orbits precess in opposing directions, the relative orbital precession rate in the frame rotating with the mode will be greater than if the mode and orbit precess in the same direction. 
However, the relative orbital precession rate will still be $\mathcal{O}(t_{\rm sec}^{-1})$, and secular torques between the orbit and the mode can still be dynamically important. 
Therefore, in near Keplerian systems, it is not dynamically forbidden for a mode to form via the Lynden-Bell mechanism with a precession direction opposite the orbital precession direction. 
Indeed, we generally see the mode precess opposite to that of the orbits in our simulations. 

In Figure~\ref{fig:mode-dynamics}, we show a 2d-histogram of $\varpi$ as a function of time for all particles in inner $a_0$ bin ($a_0 \in [100, 180]$ AU) of model N800.  We see from this figure that the individual orbits precess retrograde, and cluster together to form a mode starting around 100~$t_{\rm sec}$.  The figure shows that the mode generally precesses prograde with a pattern speed of $\sim1.5^\circ\,t_{\rm sec}^{-1}$.  The mode is capable of capturing orbits that precess counter to it.
The captured orbits then librate within the mode, precessing prograde then retrograde within the mode.
Eventually, orbits leave the mode and precess retrograde again within the disk.
\section{Simulated Observations}
\label{sec:observables}

While the inclination instability appears to be promising in explaining the clustered, detached orbits of extreme Trans-Neptunian Objects in the outer Solar System, it should also occur in exoplanet systems with at least one giant planet that can form a massive scattered disk.  In particular, it provides a mechanism for creating asymmetric debris disk structures such as the wing-like features in HD 61005 \citep{MacGregor2019}.

\begin{figure*}
    \centering
    \includegraphics[trim=2.5cm 4.5cm 3.25cm 4.25cm, clip=true, width=\textwidth]{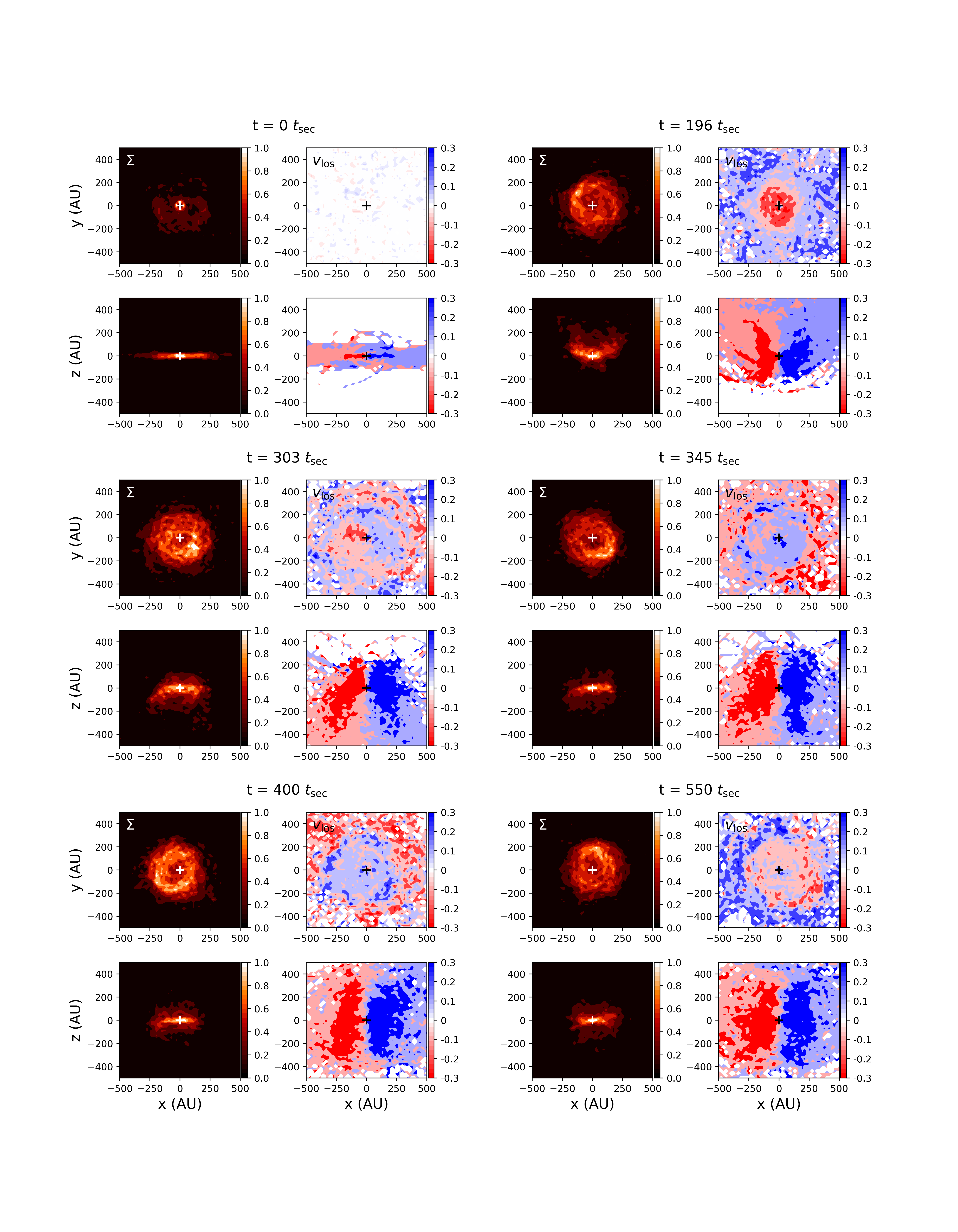}
    \caption{Snapshots in time of the J2N400 simulation\---a primordial scattered disk with the orbit-averaged gravitational influence of the giant planets.  Surface density, $\Sigma$, and velocity along the line of sight, $v_{\rm los}$ are plotted for face-on and edge-on orientations. The inclination instability occurs around 196~$t_{\rm sec}$.  At 303~$t_{\rm sec}$, we observe a spiral in the line of sight velocity when the disk is viewed face-on.}
   \label{fig:observables}
\end{figure*}

In Figure~\ref{fig:observables}, we plot snapshots of the J2N400 simulation\textemdash a primordial scattered disk with orbit-averaged gravitational influence of the giant planets. 
We first populate each orbit of the simulation with 100 particles spaced uniformly in mean anomaly to increase the effective resolution. We then make maps of surface density and velocity along the line of sight with a pixel resolution of 20 AU. 
The surface density, $\Sigma$, of the disk in face-on (top frames) and edge-on (bottom frames) orientations are plotted in the left-hand columns. Time is increasing from left to right and down the column. 
In the $x/y$-plane the particles orbit in the counter-clockwise direction. Except for the innermost edge of the disk, the orbits precess in the clockwise direction.  
The initially thin, axisymmetric disk undergoes the inclination instability, buckling above and below the plane ($t \approx$ 196\--303 $t_{\rm sec}$). The lopsided mode develops in the $x/y$-plane as differential precession disperses the asymmetric distribution of orbits in the $x/z$-plane. At early times, a spiral arm links the inner disk to the most over-dense region of the mode in the outer disk. 

To the right of the surface density plots, we plot the corresponding velocity along the line of sight, $v_{\rm los}$. Red and blue colors illustrate red-shifted and blue-shifted velocities with respect to the observer. The initial velocity distribution is dominated by rotation around the Sun, as shown in the $x/z$-plane. 
The collective rolling and pitching of the orbits about their major and minor axes (captured by the angles $i_a$ and $i_b$ in Figure~\ref{fig:sd-J2N400}) is apparent in the velocity map at $196\, \, t_{\rm sec}$ which shows resulting concentric circles of red-shifted and blue-shifted velocities. We note that this is equivalent to the clustering of the orbits in argument of pericenter ($\omega$). 

At $t \approx 303 \, t_{\rm sec}$, a spiral arm in velocity space appears in the $x/y$-plane. This occurs as the amplitude of $i_a$ for the inner orbits ($a \lesssim 500$ AU) passes through zero but their $i_b$ values are significantly non-zero, as seen in Figure 2 at t $\approx 303 \,  t_{\rm sec}$. 
The lopsided mode in the $x/y$-plane is apparent in surface density before we see the spiral arm in velocity space. 
The over-dense cluster of orbits leads to positive line of sight velocities on one side of the clump and negative on the other, leading to the appearance of a spiral. 

Another spiral arm appears in velocity space when the orbits in the disk pass through $i_a = 0$ again at $t \approx 480 \, t_{\rm sec}$ as seen in Figure 2. 
At all other times we see the concentric circles of red-shifted and blue-shifted velocities, alternating as the orbits coherently precess above and below the mid-plane. Velocities in the $x/z$-plane continue to show rotation in an increasingly thick disk.
\section{Conclusions}
\label{sec:conclusions}

In this paper, we demonstrate the spontaneous apsidal clustering of orbits of low mass bodies in $N$-body simulations of a primordial scattered disk between $\sim$100--1000 AU. 
As in \citet{zderic2020a}, we find that apsidal clustering begins after the inclination instability has saturated, and that the inclination instability is key to the formation of the lopsided mode.
In simulations where the orbit-averaged, gravitational influence of the giant planets is included, we find that apsidal clustering occurs provided that the inclination instability is not suppressed.
We also find that apsidal clustering only forms near the inner edge of the disk in the 100--320 AU range with the specific range depending on the model, but we caution that our simulations have low numbers of particles particularly at large semi-major axes.
The fast orbital precession caused by the giant planets pushes the location of apsidal clustering out to larger semi-major axis.
Finally, we find that the resulting lopsided mode strength oscillates, but appears long-lasting.

\citet{lyndenbell1979} proposed a mechanism to explain stellar bar formation in the center of galaxies that we extend here to near-Keplerian systems to explain the apsidal clustering that occurs in our simulations. 
Orbit-averaged torques from a weak, lopsided mode encourages orbits into precessing towards alignment with the mode.\footnote{We assume an initial small mode is seeded by random fluctuations within the disk.} 
In a Keplerian system, if $\nicefrac{\partial{\dot{\varpi}}}{\partial{e}} |_{a} < 0$ then orbits will tend to align with and reinforce the mode.
We call regions of $e$-$a$ space where $\nicefrac{\partial{\dot{\varpi}}}{\partial{e}} |_{a} < 0$ clustering regions.

We have created contour plots of $\dot{\varpi}$ as a function of eccentricity and semi-major axis within the disk at different times.
We find that a clustering region forms during the peak of the inclination instability when the disk has formed a bowl-shape.
The clustering region appears just before apsidal clustering begins, and it appears at the inner edges of the disk. 
In simulations with the orbit-averaged gravitational influence of the giant planets, the added $J_2$ inhibits circularization of the inner edge of the disk during the instability and amplifies circularization at larger semi-major axis.
As a result, the clustering region is populated by bodies with larger semi-major axis in the $J_2$ models and apsidal clustering correspondingly occurs at larger semi-major axes.

\begin{figure*}[!htb]
    \centering
    \includegraphics{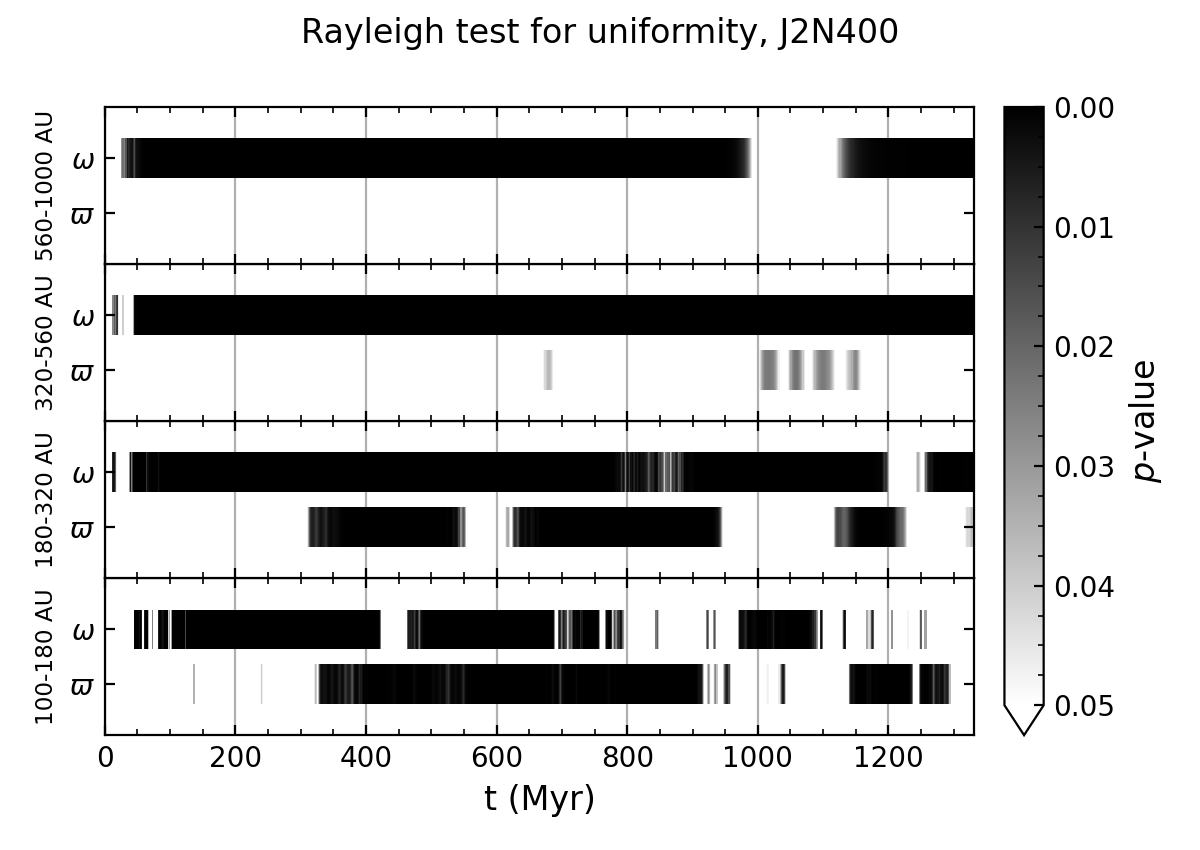}
    \caption{Rayleigh test results for $\omega$ and $\varpi$ binned by initial semi-major axis as a function of time for the J2N400 model scaled such that the inclination instability saturates (ceases exponential growth) at 250 Myr. Rayleigh test $p$-values less than 0.05 indicate that the distribution of $\omega$ or $\varpi$ is not uniform (i.e. it is clustered) and smaller $p$-values indicate stronger clustering. $\omega$-clustering begins after $\sim$10 Myr and persists until the end of the simulation at $\sim$1300 Myr, except for bodies with $a_0\in[100,180]$~AU. $\varpi$-clustering occurs for bodies with $a_0\in[100,320]$~AU after the inclination instability has saturated at $\sim$250 Myr, and it disappears and reappears intermittently.}
    \label{fig:clustering-J2N400}
\end{figure*}
\begin{figure*}[!htb]
    \centering
    \includegraphics{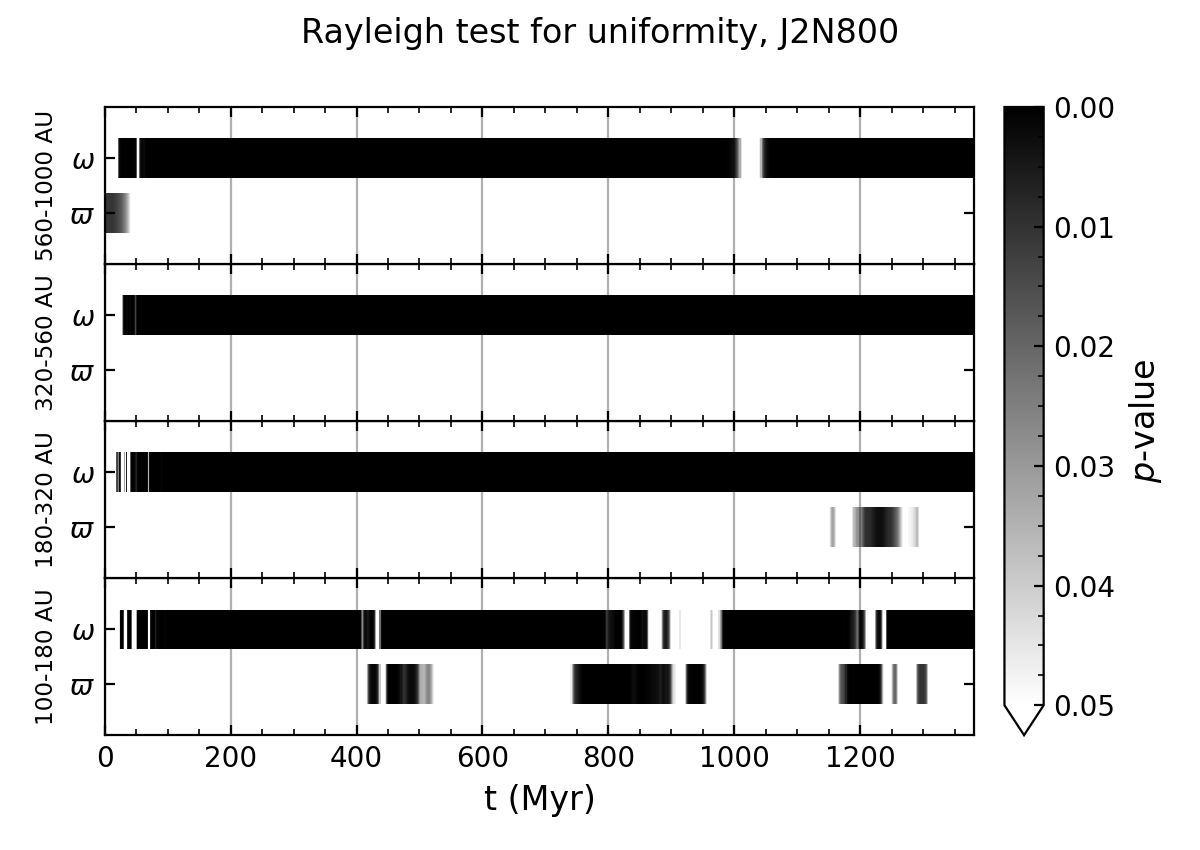}
    \caption{Rayleigh test results for $\omega$ and $\varpi$ binned by initial semi-major axis as a function of time for the J2N800 model scaled such that the inclination instability saturates (ceases exponential growth) at 250 Myr. Rayleigh test $p$-values less than 0.05 indicate that the distribution of $\omega$ or $\varpi$ is not uniform (i.e. it is clustered) and smaller $p$-values indicate stronger clustering. $\omega$-clustering begins after $\sim$10 Myr and persists until the end of the simulation at $\sim$1300 Myr. $\varpi$-clustering occurs primarily in bodies with $a_0\in[100,180]$~AU after the inclination instability has saturated, and it is more intermittent than in the J2N400 model. The initial $\varpi$-clustering in the outer $a_0$ bin is a random quirk of our initial conditions, and it quickly disappears.}
    \label{fig:clustering-J2N800}
\end{figure*}

The clustering region is directly correlated with the unique bowl-shaped orbital distribution created by the inclination instability. 
Due to orbital precession, the bowl-shaped distribution oscillates back and forth across the original plane of the disk, causing the clustering region to repeatedly disappear and reappear, and eventually, the bowl-shape disappears.
However, we find that the lopsided mode created by the clustering region persists.
We hypothesize that the mode eventually becomes massive enough to trap orbits without the help of the background disk potential. 

Surface density plots of our disks during the inclination instability show edge-on wing-like structures reminiscent of some debris disks (e.g. HD61005), and a lopsided mode in face-on views after the instability has saturated. 
In line-of-sight velocity, we see concentric circles of alternating sign associated with the bowl-shaped orbital distribution post-instability. 
Later, the lopsided mode creates spiral arms in line-of-sight velocity. 
Observational signatures like this in exoplanet disks could be caused by the inclination instability provided there is something to pump-up the orbital eccentricity of the bodies in the disk (e.g. a giant planet).

In \citet{Zderic2020b}, we found that $\sim$20 Earth masses is required for a primordial scattered disk to resist the orbit-averaged quadrupole potential of the giant planets \textit {at their current locations} and undergo the inclination instability\footnote{Twenty earth masses is extreme for a primordial scattered disk (indeed perhaps too massive for the instability to have occurred in our solar system). While hundreds of earth masses may have been present in the early planetesimal disk beyond $\sim5$ AU, only a fraction appears to pass through the scattered disk region \citep{Nesvorny2018}.}.

The e-folding timescale for the inclination instability in a scattered disk configuration with $N\rightarrow\infty$ and without added $J_2$ is \citep{Zderic2020b},
\begin{equation}
    t_{\rm e-fold} \sim \frac{2.4}{\pi} \frac{M_\odot}{M_d} P.
\end{equation}
For a 20 Earth mass disk, $t_{\rm e-fold} = 1.3\times10^{4}\,P$. 
Based on \citet{Zderic2020b} Figure 3, we expect this timescale to be increased by a factor of $\sim$4 in simulations with added $J_2 \lesssim J_{2,{\rm crit}}$.
With the inner edge semi-major axis being $100\,{\rm AU}$, $P = 1000\,{\rm yr}$,
and the e-folding timescale for the inclination instability in a 20 Earth mass Scattered Disk in the outer solar system is $\sim$50 Myr.
It takes about five e-folding timescales for the inclination instability to saturate. Therefore, the inclination instability in this primordial scattered disk should saturate after $\sim$250 Myr.

We can estimate the duration of $\omega$ and $\varpi$ clustering in the 20 Earth mass primordial scattered disk using the J2N400 and J2N800 simulations. We set the saturation time in these simulations to be 250 Myr and scale the subsequent evolution of the disk using the secular timescale. For example, $t_{\rm sec} =  2.6\,{\rm Myr}$ (see Equation~\ref{eq:tsec}) for a 20 Earth mass disk. The J2N800 simulations runs for $\sim$400 $t_{\rm sec} \approx 1050\,{\rm Myr}$ after the saturation of the instability.
We stress that this scaling is approximate and meant to provide a qualitative, order-of-magnitude estimate for the duration of angular clustering in an unstable primordial scattered disk.
In Figures~\ref{fig:clustering-J2N400} and \ref{fig:clustering-J2N800}, we show the $p$-values for the Rayleigh $z$ test for uniformity \citep{MardiaAndJupp} on $\omega$ and $\varpi$ for the J2N400 and J2N800 models as a function of time in Myr using this proposed scaling. 
Rayleigh $z$ test $p$-values less than 0.05 signify that the angular distribution is not consistent with a uniform distribution, i.e. that the distribution is clustered.
We chose the Rayleigh test as it is sensitive to unimodal deviations from uniformity.
In both models, $\omega$-clustering begins after just a few tens of Myr and persists for the duration of the simulation except in the inner semi-major axis bin (100-180 AU) where differential precession is the strongest.
Intermittent clustering in $\varpi$ begins after the inclination instability has saturated in the inner two bins in both simulations.
Note that binning the results by $a_0$ partially mitigates the effects of differential precession, prolonging the duration of angular clustering. 
Overall, we expect a 20 Earth mass primordial scattered disk to be able to sustain $\omega$-clustering for $a \gtrsim 180$~AU (if binned by semi-major axis) and intermittent periods of $\varpi$-clustering for $a\in[100,320]$~AU for Gyr timescales.
    
The inclination instability can raise perihelia and inclinations of bodies in the outer solar system. As such, it can effectively trap planetesimal mass at semi-major axes of hundreds of AU as bodies are isolated from strong scattering encounters with the giant planets. In \citet{Zderic2020b} we show that orbits with semi-major axes between $\sim 200-500$ AU end up with a (rather extraordinary) median perihelion distance of 150 AU post-instability; see e.g., Figure 6. The observed Sednoids in this scenario mark the inner edge of a massive reservoir of extremely detached bodies originating from the primordial scattered disk. 

The mass remaining at hundreds of AU today is an open question that we are actively exploring. Our simulations show that, post-instability, inter-orbit torques induce eccentricity (but not necessarily inclination) oscillations on particles in this structure. In future work we will calculate the flux of particles back into the inner solar system through these oscillations (which can cause perihelia to drop below the orbit of Neptune) and ultimately the mass loss rate. 

\acknowledgments%%
\section*{Acknowledgements}

We thank the referees for their insightful comments and suggested changes which improved the quality of our manuscript. AM gratefully acknowledges support from the David and Lucile Packard Foundation.
This work was supported by a NASA Solar System Workings grant (80NSSC17K0720) and a NASA Earth and Space Science Fellowship (80NSSC18K1264). 
This work utilized resources from the University of Colorado Boulder Research Computing Group, which is supported by the National Science Foundation (awards ACI-1532235 and ACI-1532236), the University of Colorado Boulder, and Colorado State University. 

\software{\texttt{REBOUND} \citep{Rein2012}, 
\texttt{REBOUNDX} \citep{Tamayo2019}, \texttt{GNU Parallel} \citep{Tange2011a}}

\appendix
\section{Finding the Clustering Region}
\label{app:finding-clustering}

\begin{figure*}[!htb]
	\centering
 	\includegraphics[]{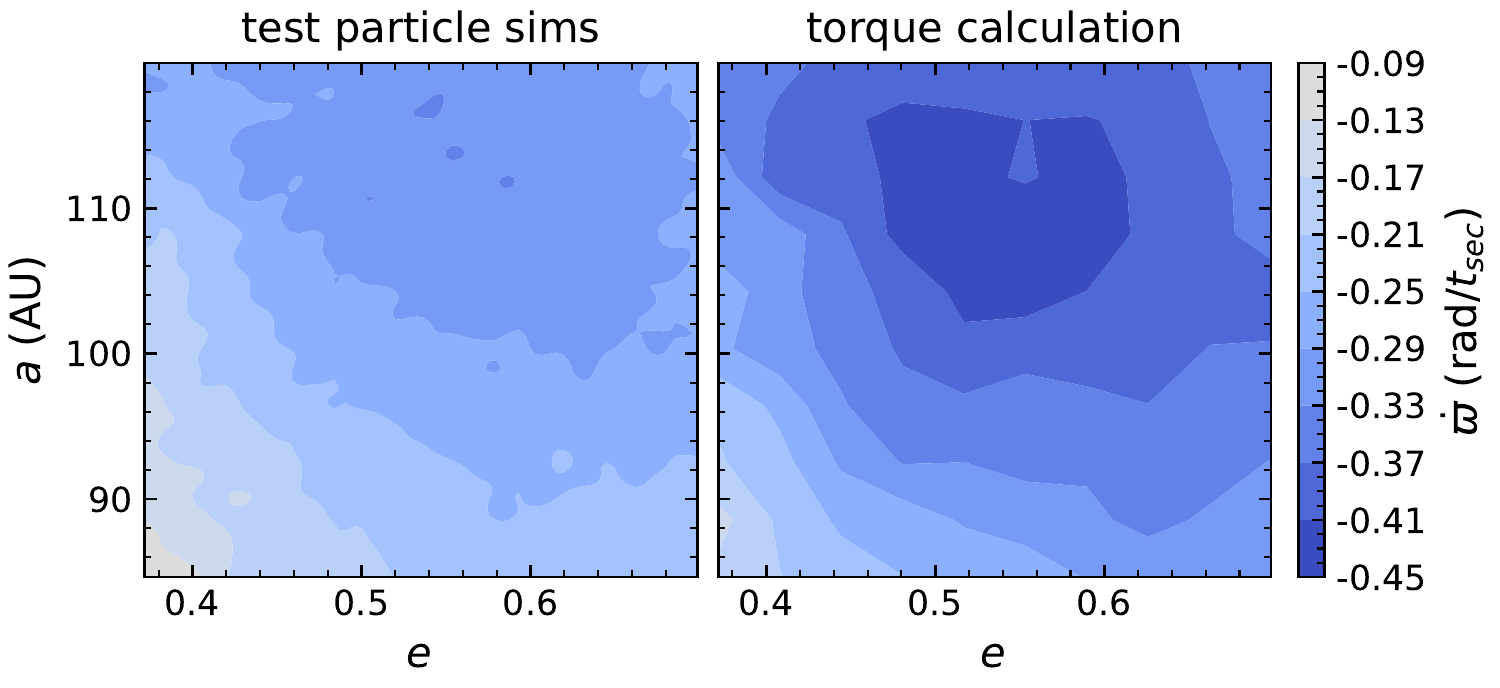}
	\caption{Comparison of the two methods used to calculate the precession rate in the frozen disk potential. The left panel shows $d\varpi/dt$ calculated from a test particle simulation on a 100 by 100 $e$-$a$ grid while the right panel shows $d\varpi/dt$ calculated with the torque method from Appendix~\ref{app:finding-clustering} on a 10 x 10 grid. The two methods agree qualitatively, though the torque calculation gives faster precession rates than the test particle simulations. Both plots were made using data from an $N=400$ compact configuration simulation without added $J_2$.}
	\label{fig:method-compare}
\end{figure*}

In Section~\ref{sec:clustering-region-emerges}, we show the clustering region within the disk using contour plots of $\dot{\varpi}$ in $e$-$a$ space. 
In these plots, $\dot{\varpi}$ is calculated using test particle simulations described in the same section.
Alternatively, we can calculate instantaneous precession rate of a test orbit can be directly from the torques and forces it experiences without integrating the orbit. In particular, the orbital precession rate can be computed from the torque and the time derivative of the eccentricity vector, viz.
\begin{align}
    \dot{\bm{e}}=\frac{\bm{f \times j}}{G M}+\frac{\bm{v \times \tau}}{G M},
    \label{eq:ederiv}
\end{align}
where $\bm{f}$ and $\bm{\tau}$ are the specific force and torque on the test orbit; $\bm{v}$ is the velocity;  $M$ is the central mass (see equation 1 in \citealt{madigan2017} and surrounding discussion).

In order to validate the results of \S~\ref{sec:clustering-region-emerges}, we use equation~\eqref{eq:ederiv} to calculate the precession rates of test orbits injected into $N$-body simulations. Specifically, we
\begin{enumerate}
    \item Discretize each orbit into one thousand, equal mass points, evenly-spaced in mean anomaly.
    \item Compute the total force and torque ($\bm{\tau}$) on each point along the test orbit from all of the disk orbits.
    \item Use equation~\eqref{eq:ederiv} to determine $\dot{\bm{e}}$ at each point along the test orbit. 
    \item Average $\bm{\tau}$ and $\dot{\bm{e}}$ over the test orbit.
    \item Check for convergence by repeating the above steps with a new set of discretrized point (evenly spaced in mean anomaly between existing ones). With one thousand points, the results are always converged within 10\%. (And usually it is much better: for 90\% of test orbits we obtain convergence within 3\%.)
\end{enumerate}
Similar methods have previously been used in studies of resonant relaxation in the Galactic center (e.g. \citealt{gurkan&hopman2007}). 
To compare to the precession rate in the preceding section it is necessary to convert from $\dot{\bm{e}}$ and $\bm{\tau}$ to $\dot{\varpi}=\dot{\Omega}+\dot{\omega}$.\footnote{Note that $\dot{\varpi}=\dot{\Omega}-\dot{\omega}$ for retrograde orbits, which are not considered here.} The Kepler orbital angles, $\Omega$ and $\omega$, are related to the eccentricity and angular momentum vectors as follows
\begin{align}
    &\Omega=\arctan\left(\frac{\hat{n}_y}{\hat{n}_x}\right)\nonumber,\\
    &\omega=\arccos\left(\hat{n}\cdot \hat{e}\right)\nonumber,\\
    &\bm{n}=\hat{\textbf{z}} \bm{\times} \hat{\textbf{\j}}.
\end{align}
Then $\dot{\Omega}$ and $\dot{\omega}$ can be approximated as 
\begin{align}
    &\dot{\Omega} \approx \frac{\Omega(\bm{j} (t+\delta t), \bm{e} (t+\delta t))-\Omega(\bm{j} (t), \bm{e} (t))}{\delta t}\nonumber,\\
    &\dot{\omega} \approx \frac{\omega(\bm{j} (t+\delta t), \bm{e} (t+\delta t))-\omega(\bm{j} (t), \bm{e} (t))}{\delta t}\nonumber,\\
    &\bm{j} (t+\delta t)\approx \bm{j}(t)+ \bm{\tau}\delta t\nonumber,\\
    &\bm{e} (t+\delta t)\approx \bm{e}(t)+ \bm{e'}\delta t,
\end{align}
where $\delta t$ is a small time interval. Here we use $\delta t=10^{-6} |\bm{j}|/|\bm{\tau}|$. We have verified that the results do not depend on $\delta t$.

The precession rate calculated using this torque method is compared to the precession rate from the test particle simulations in Figure~\ref{fig:method-compare}.
Note that this plot shows results from a compact orbital configuration not a scattered disk orbital configuration. 
The compact configuration is an axisymmetric, nearly-flat disk of Keplerian orbits in which all bodies have identical eccentricities and nearly identical semi-major axes. 
This limited radial structure simplifies analysis. 
The two methods qualitatively agree, however, the test particle simulation method gives slower precession rates than the torque calculation. This is because two-body interactions in the test particle simulations (not accounted for in the torque calculation) weaken secular torques.

\section{Measuring apsidal clustering}
\label{app:measure-apsidal}

\begin{figure*}[!thb]
	\centering
	\includegraphics[width=0.75\textwidth]{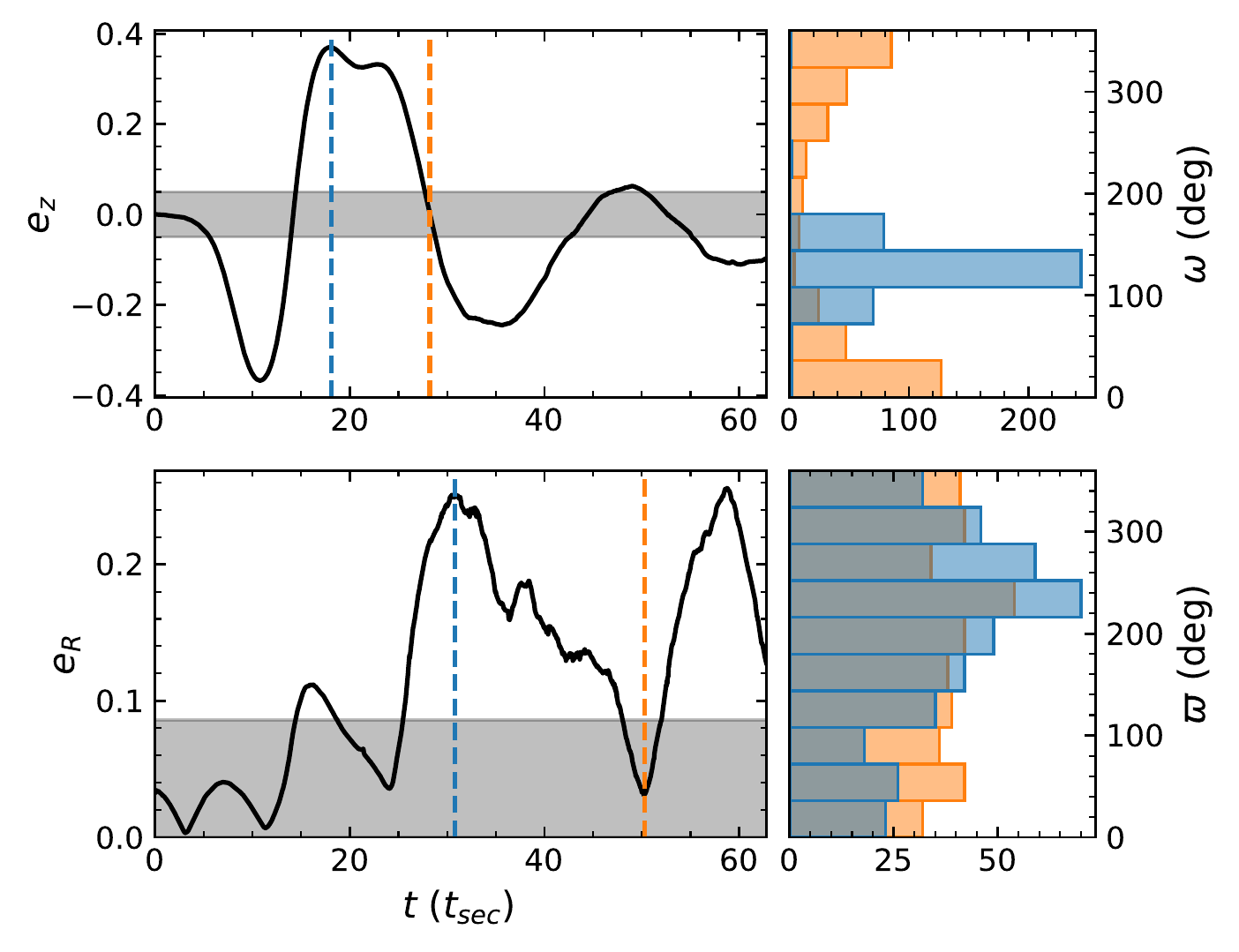}
	\caption{Post-instability apsidal clustering in a compact orbital configuration. The top row shows $e_z$ on the left and a histogram of each particle's $\omega$ on the right at the two times marked by colored vertical lines in the $e_z$ plot. The bottom row shows $e_R$ on the left and a histogram of each particle's $\varpi$ on the right at the two times marked in the $e_R$ plot. This shows that the peaks in $e_R$ correspond to $\varpi$ clustering, and that zero $e_z$ can still be $\omega$-clustered. The inclination instability saturates at around $10\,t_{\rm sec}$ in this simulation. This data comes from a 400 particle compact configuration simulation without added $J_2$.}
	\label{fig:compact-clustering}
\end{figure*}
\begin{figure*}[!thb]
	\centering
	\includegraphics[width=0.75\textwidth]{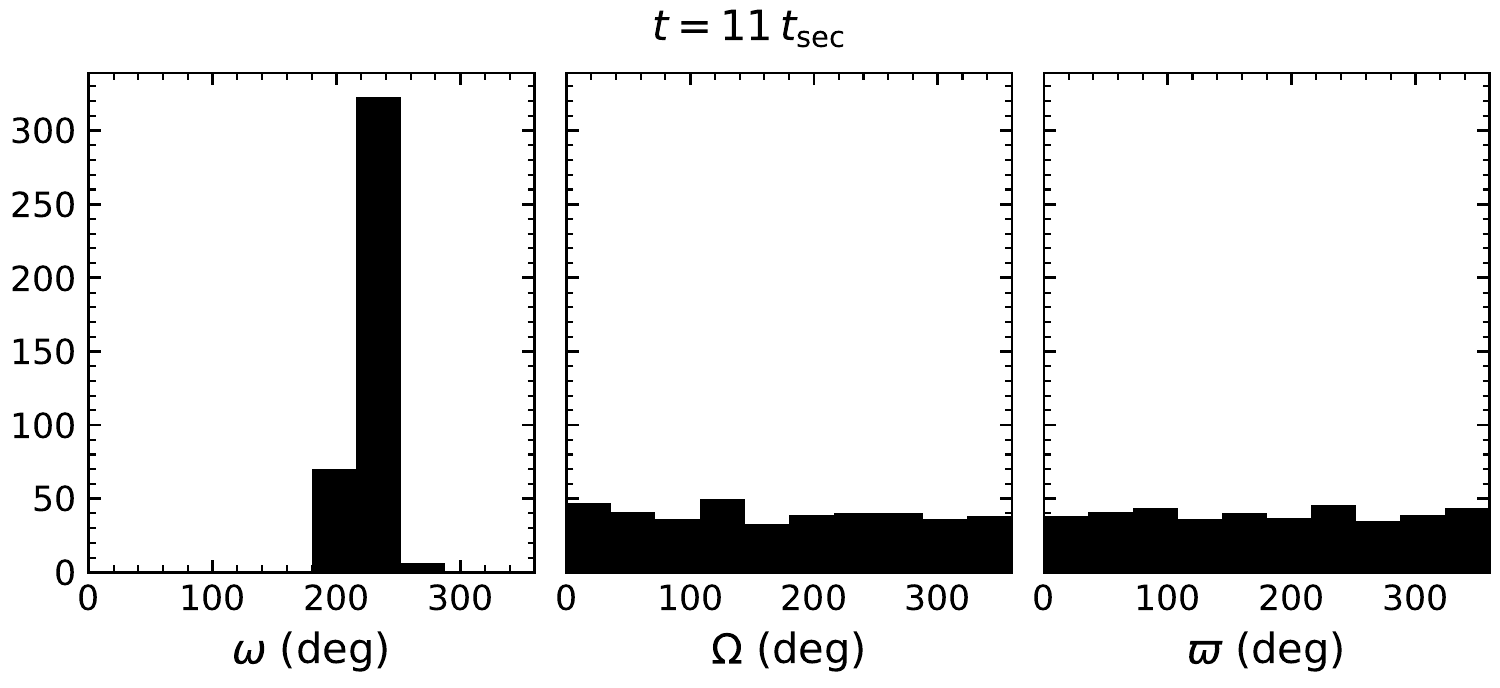}
	\caption{Orbital angles of bodies from the compact configuration simulation shown in Figure~\ref{fig:compact-clustering} at $t=11\,t_{\rm sec}$. Arguments of pericenter $\omega$ are highly clustered, while $\Omega$ and $\varpi$ show no clustering and are statistically consistent with a uniform distribution (Kuiper's test). This is what the bowl-shape driven by the inclination instability looks like in Kepler angles.
	}
	\label{fig:aop-clustering}
\end{figure*}

Here we demonstrate the connection between the components of the mean, normed eccentricity vector, $e_R$ and $e_z$ (see equation~\ref{eq:eReTez}), and the Kepler elements argument of pericenter, $\omega$, and longitude of pericenter, $\varpi$ .
In Figure~\ref{fig:compact-clustering}, we reproduce a result from \citet{zderic2020a} in which we demonstrate the appearance of apsidal clustering in a simulation of particles in a compact orbital configuration.
Initially, the disk is axisymmetric; $e_R$ is below the noise floor. As the top panel shows, the inclination instability begins at $t \lesssim t_{\rm sec}$ and saturates at $\sim\!10\,t_{\rm sec}$. After the instability saturates and orbits apsidally precess back through the mid-plane ($e_z \approx 0$), we begin to see statistically significant $e_R$, indicating in-plane apsidal clustering. The right panels show histograms of $\omega$ and $\varpi$ of all bodies at two times which are marked using colored-matching dashed lines. An $e_R$ above the noise floor corresponds to $\varpi$-clustering but an $e_z$ below the noise floor can still be $\omega$-clustered.

Using the mean unit eccentricity vector instead of $\varpi$ to measure apsidal clustering has two advantages: the mean unit eccentricity vector is 3D and can capture out-of-plane clustering, and statistical analyses on compound angles like $\varpi$ can be misleading. 
We demonstrate the first point in Figure~\ref{fig:aop-clustering}. At $t\sim10\,t_{\rm sec}$, the bodies in the disk have a uniform $\varpi$ distribution suggesting that there is no apsidal clustering. 
However, the $z$ component of the mean unit eccentricity vector is large (see top left panel of Figure~\ref{fig:compact-clustering}), indicating that the orbits apses are strongly clustered perpendicular to the plane. 

Statistics on a compound angle can be misleading. If either $\omega$ or $\Omega$ is uniformly distributed, and $\omega$ and $\Omega$ are independent and have continuous distributions, then $\varpi$ will also be uniformly distributed. 
In essence, $\omega$ or $\Omega$, whichever is uniformly distributed, has the capacity to erase the others distribution in $\varpi$.
This can be seen in Figure~\ref{fig:aop-clustering}.
At $11\,t_{\rm sec}$, $\omega$ is highly clustered, nearly a delta function, while both $\Omega$ and $\varpi$ are uniformly distributed.

We now prove this. We define the normalized distributions of $\varpi$, $\omega$, and $\Omega$ as $f(\varpi)$, $g(\omega)$, and $h(\Omega)$, and recall that $\varpi = \omega + \Omega$.
These distributions are periodic, e.g., $f(\varpi) = f(\varpi+2\pi)$.
The distribution of the sum of two continuous, independent random variables is given by the convolution of the two distributions,
\begin{equation}
    f(\varpi) = \int_0^{2\pi} g\left(\varpi-\Omega\right) h\left(\Omega\right) d\Omega.
\end{equation}
If the distribution of $\Omega$ is uniform, $h(\Omega) = \nicefrac{1}{2\pi}$, then,
\begin{equation}
    f(\varpi) = \frac{1}{2\pi} \int_0^{2\pi} g\left(\varpi-\Omega\right)d\Omega.
\end{equation}
Switching back to $\omega$,
\begin{align}
    f(\varpi) &= \frac{1}{2\pi} \int_{\varpi-2\pi}^{\varpi} g(\omega)d\omega, \\
              &= \frac{1}{2\pi} \int_{\varpi}^{\varpi + 2\pi} g(\omega - 2\pi)d\omega, \\
              &= \frac{1}{2\pi} \int_{\varpi}^{\varpi + 2\pi} g(\omega)d\omega, \\
              &= \frac{1}{2\pi} \int_{0}^{2\pi} g(\omega)d\omega, \\
              &= \frac{1}{2\pi},
\end{align}
where we've used the normalization of $g(\omega)$,
\begin{equation*}
    1 = \int_{0}^{2\pi} g(\omega) d\omega,
\end{equation*}
and the identity,
\begin{equation*}
    \int_{y}^{y + 2\pi} F(x)dx = \int_{0}^{2\pi} F(x)dx,
\end{equation*}
which holds for any $y\in\mathbb{R}$ and function $F(x)$ periodic with period $2\pi$.
This proof holds in the case where $\omega$ and/or $\Omega$ is uniformly distributed.

%%%%%%%%%%%%%%%%%%%%%%%%%%%%%%%

\footnotesize{
\bibliographystyle{aastex}
\bibliography{main.bib}
}

%%%%%%%%%%%%%%%%%%%%%%%%%%%%%%%

\end{document}